# Deployment and calibration procedures for accurate timing and directional reconstruction of EAS particle-fronts with HELYCON stations

T. Avgitas[ae], G. Bourlis[a], G.K. Fanourakis[b], I. Gkialas[c], A. Leisos[a], I. Manthos[c], A. Tsirigotis[a], S.E. Tzamarias[ad†]

a *Physics Laboratory, School of Science & Technology, Hellenic Open University Patras, Greece*

b *Institute of Nuclear and Particle Physics, NCSR Demokritos, Athens, Greece*

c *Department of Financial and Management Engineering, University of the Aegean, Chios, Greece*

d *Department of Physics, Aristotle University of Thessaloniki, Thessaloniki, Greece*

e *Laboratory APC, University Paris Diderot - Paris VII, Paris, France*

## Abstract

High energy cosmic rays, with energies thousands of times higher than those encountered in particle accelerators, offer scientists the means of investigating the elementary properties of matter. In order to detect high energy cosmic rays, new detection hardware and experimental methods are being developed. In this work, we describe the network of HELYCON (HEllenic LYceum Cosmic Observatories Network) autonomous stations for the detection and directional reconstruction of Extended Atmospheric Showers (EAS) particle-fronts. HELYCON stations are hybrid stations consisting of three large plastic scintillators plus a CODALEMA antenna for the RF detection of EAS particle-fronts. We present the installation, operation and calibration of three HELYCON stations and the electronic components for the remote control, monitor and Data Acquisition. We report on the software package developed for the detailed simulation of the detectors' response and for the stations' operation. The simulation parameters have been fine tuned in order to accurately describe each individual detector's characteristics and the operation of each HELYCON station. Finally, the evaluation of the stations' performance in reconstructing the direction of the EAS particle-front is being presented.

## Introduction

We have previously reported [1] on the construction, tests and performance evaluation of large charged particle detectors (plastic scintillator counters) to be used in detecting Extensive Atmospheric Showers (EAS) in a network of independent stations. In this paper, we report on the installation and operation of three such stations, deployed at the Hellenic Open University (HOU) campus. Each of these stations comprises three large charged particle detectors and one RF detection system (CODALEMA antennas) [2], equipped with trigger, digitization and Data Acquisition (DAQ) electronics. It is also equipped with slow control and monitor electronics and a GPS-based timing system, whilst the DAQ and the control/monitor online-software packages are hosted on a Station Local Computer (SLC). The SLCs are connected, via wireless Ethernet, to the Global Data and Control Server (GDCS), where the data are transferred. The operation and control of the stations are performed remotely through the GDCS. Furthermore, a fourth station (excluding the antenna) has also been installed in the premises of the HOU Physics Laboratory, 4 km apart from the other stations, which is used for both data acquisition, as well as for special calibration and operational tests.

† Corresponding author

Standard installation, calibration and operation procedures have been established and software for controlling, monitoring and collecting data from the detectors' network has been developed. Furthermore, software packages to simulate in detail the response of the detectors to particle showers as well as to reconstruct the EAS characteristics have been advanced. The detector network has been operated, collecting data for more than a year and a half. Its performance in detecting and reconstructing EAS has been evaluated by comparing with the simulation predictions. Despite the limited size, this detector array offered a workbench for the development of tools and techniques related to EAS detection [3, 4, 5]. A larger network, HELYCON (Hellenic Lyceum Cosmic Observatories Network), is under construction following the procedures, the techniques and the software tools, which are described in the following sections.

This paper reports on the new developments mentioned above and on results concerning the performance of the deployed detector network. Section 1 describes briefly the large charged particle detectors (henceforth called HELYCON detectors) and their arrangement in the deployed HELYCON stations. It also includes the description of the control/monitor and the Data Acquisition systems. Section 2 focuses on the installation procedures utilized for the particle detectors of each HELYCON station, in order to calibrate and synchronize the detectors. It starts with an outline of the Monte Carlo package, which is used to simulate the particle detector response. It reports on the tuning of the simulation parameters in order to precisely describe each individual particle counter as well as the functioning of each HELYCON station. Section 3 concentrates to the evaluation of the stations performance to reconstruct the direction of the detected EAS. Conclusions are presented in the final Section.

Different papers (to be published) report on results based on the analysis of the collected data from each station, as well as of the data from more than one station that correspond to the same EAS, and on the analysis of data recorded synchronously by both the large particle detectors and the CODALEMA RF antennas of a HELYCON station.

## 1. The Deployed HELYCON Array

The three stations, which have been installed at the Hellenic Open University campus in Patras, follow the arrangement shown in Figure 1. The inter-station distances are of the order of a few hundred meters (i.e. 164 m between Station-1 and Station-2, 467 m between Station-2 and Station-3, 328 m between Station-3 and Station-1). Another station, Station-4, has been installed in the Physics Laboratory (coordinates 38°14'40.56"N, 21°43'47,55"E), about 4 km apart, which is used for special calibration purposes. Each station consists of the 3 HELYCON detectors positioned around 30 m apart and one CODALEMA RF detection system, placed among the particle detectors. Their relative positions are as it is shown in Table 1. Due to the fact that new HELYCON stations will be deployed on the top of school buildings, most of them offering an elongated space, Station-2 has been chosen to be installed in a similar, asymmetric arrangement, in order to study the efficiency and the geometrical acceptance of HELYCON stations in such geometries. Furthermore, in order to study external electronic noise rejection techniques, in cases where stations will be installed close to noise producing devices, Station-2 is positioned next to the central air-conditioning system of the HOU campus. The electronics for powering, controlling and monitoring the particle detectors and the RF system are housed in a special metallic, waterproof box, henceforth called Station Electronics Box (SEB). SEB also houses the Digitization and Data

Acquisition (DAQ) electronics for the particle detectors and the station's local computer (SLC). The SLC provides the means for remote control and monitoring of the whole station and also hosts the control/monitoring and the DAQ software for the particle detectors. All the control, signal and power cables, connecting the detectors with the SEB are protected within plastic tubes, buried under the ground where required. Station-4 is similar to the other stations but does not include any RF system (mainly due to the high ambient electronic noise).

|  | Detector 1 | | | Detector 2 | | | Detector 3 | | |
| --- | --- | --- | --- | --- | --- | --- | --- | --- | --- |
|  | x(m) | y(m) | z(m) | x(m) | y(m) | z(m) | x(m) | y(m) | z(m) |
| Station-1 | 10.73 | 6.23 | 0.45 | 4.97 | -14.13 | 0.29 | -17.88 | 4.42 | 0.00 |
| Station-2 | 26.10 | 1.03 | 0.62 | -14.37 | 1.61 | 0.32 | 5.68 | -5.49 | 0.00 |
| Station-3 | -0.18 | 7.46 | 0.00 | 12.48 | -9.61 | 0.00 | -16.58 | -7.82 | 0.00 |
| Station-4 | 12.67 | -5.16 | 0.00 | -4.52 | -4.26 | 0.85 | 5.70 | 2.24 | 0.27 |

Table 1: Relative charged particle detectors' positions within stations. x points to the east, y to the north and z is the altitude. The coordinate system is centered at the antenna of each station.

The HELYCON stations operate independently of each other, under the supervision of their SLC. External communication with any station is only possible through the central server, GDCS, of the HELYCON array by connecting through wireless Ethernet to the appropriate SLC.

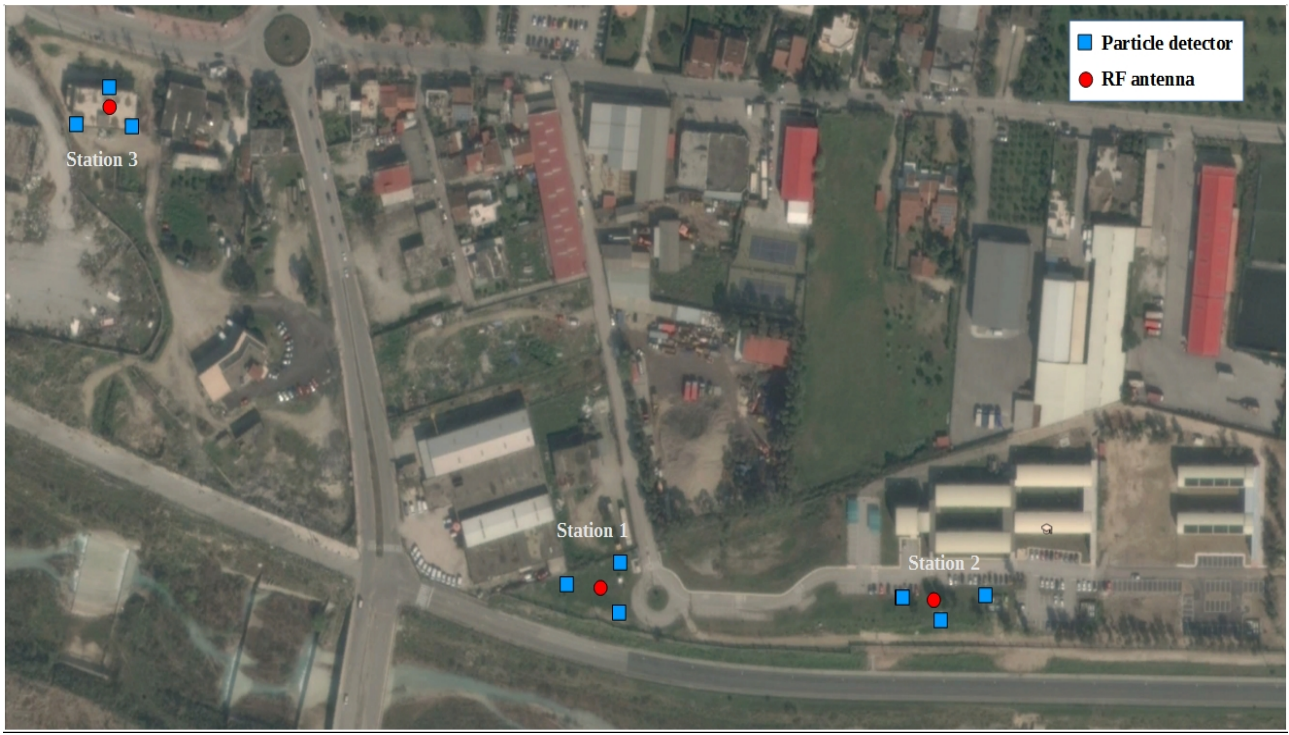

Figure 1: Layout of the HELYCON array installed at the Hellenic Open University campus. The geodesic coordinates of the stations centers are as follows: Station 1 (38°12'22.69"N, 21°45'52.60"E), Station 2 (38°12'22.49"N, 21°45'59.32"E), Station 3 (38°12'29.62"N, 21°45'42.36"E).

The RF detection system of the HELYCON stations is the CODALEMA system [2], triggered by the response of the charged particle detectors of each station, as described in Section 1.2. Apart from the external triggering, the CODALEMA system is independent, and operates by means of its own, signal detection, filtering, time-tagging of the events, DAQ and storage modules, whilst also possesses the ability of self-triggering. A standard UTP cable connects the PC board of the RF system with the station's SLC and this connection is used to set operation parameters and controls, as well as to transfer the collected data to the central server, GDCS, of the HELYCON Array.

Due to the common trigger for each selected event, information is recorded by both DAQs of the particle detectors and the RF system of the station. The two information pieces are time-tagged using absolute timing provided by GPS devices embedded in the corresponding DAQs. The event building, i.e. the combination of the experimental information provided by the RF system and the particle detectors, is performed offline, based on the time-tags of the events. Similarly, the formation of global-events, when an EAS has been recorded by two or more stations, is based on this time information and it is also performed offline.

The operation of the RF system within the HELYCON array, including data collection, signal processing and data analysis, as well as the detailed description of the event building procedure and the analysis of corresponding global events are presented in another paper (published in parallel).

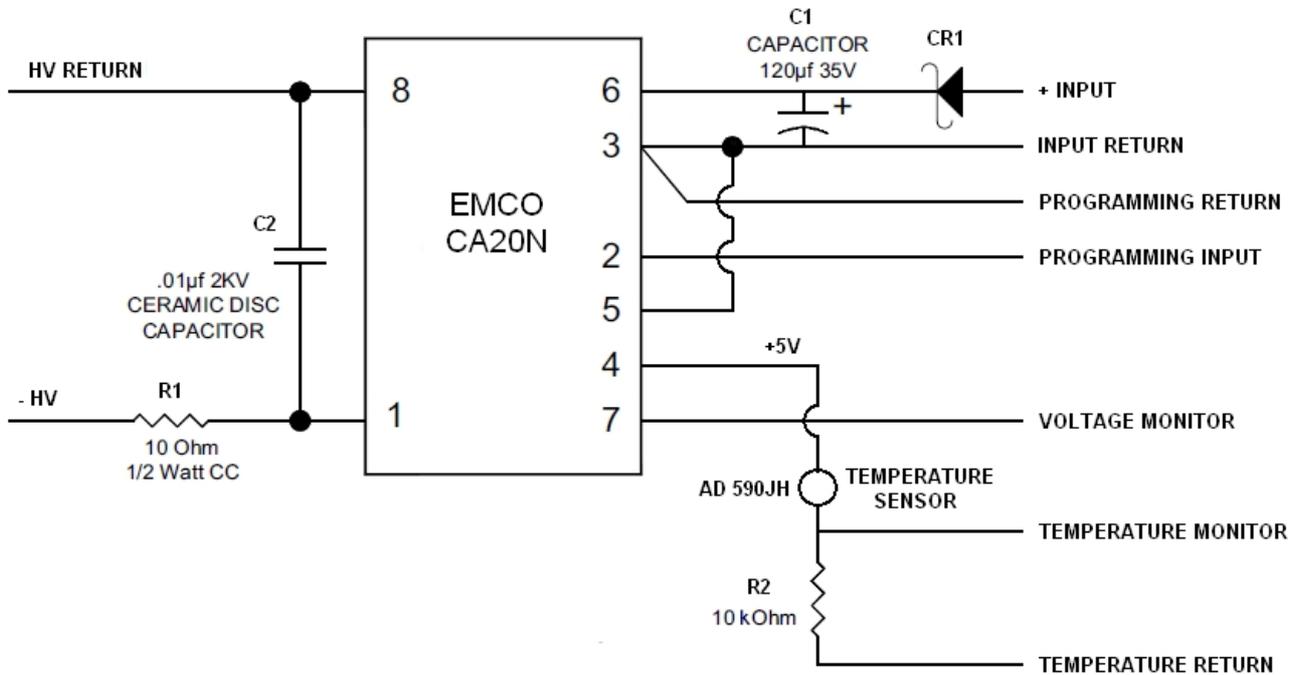

Figure 2: Schematic of the electronic board that provides and monitors the high voltage of the PMTs of the HELYCON detectors. The board includes a temperature sensor, monitoring the temperature next to the electronics of the detector.

### 1.1. The HELYCON charged particle detectors

HELYCON charged particle detectors [1, 6] are 1 $m^2$ plastic scintillator counters, made of two layers of 80 (12x10x0.5 $cm^3$) scintillator tiles per layer, wrapped in reflective paper (Dupont Tyvek 2460B) and adjusted on a wooden frame. The scintillation material is IHEP SC-301 [7] containing 2% pTP (p-terphenil) and 0.02% POPOP dopants. The light, which is generated by the interaction of particles with the scintillation material, is collected by wavelength shifting optical fibers (Bicron BCF91-A [8]) that are embedded inside grooves build in the scintillator tiles and is driven to a fast Photonis XP1912 [9] photomultiplier (PMT) for detection. The high voltage required by the PMT is provided by means of an electronic board based on the remotely controlled high voltage DC-DC converter EMCO CA20N [10], powered externally with 12 V. This board also includes a temperature sensor as it is depicted in Figure 2. The PMT, the temperature sensor and the electronic board are adjusted on the wooden frame of the detector. The charged particle detector (i.e. the scintillator with the optical fibers, the PMT, the thermometer and the electronics board) is shielded against the ambient electronic noise by means of a thick, 0.4 mm, aluminum foil that is wrapped around its wooden frame. Each HELYCON counter was set inside a water-proof wooden container, made out of boat-building wood, which consequently was covered with reflective insulator (Tyvek), to prevent overheating and excessive humidity inside the detector.

To control and monitor the high voltage supply, as well as to monitor the environmental temperature around the detector, an external USB Digital/Analog I/O device (National Instruments NI-USB 6008 [11], hereafter called I/O device) is employed to provide and record the necessary signals. Each station requires two such devices, which are hosted inside the SEB of the station and are controlled via the SLC. One 12 V DC power supply, housed inside the SEB, is needed to provide power to the three charged particle detectors of each station. Connections between the I/O devices and the electronic boards of the detectors are achieved using standard UTP cables. Coaxial RG58 cables are used for transferring the detector signals to the digitization and DAQ electronics (that is a Quarknet board [12], housed inside the SEB), as well as to provide power to the detectors and a common ground to all the electronic devices of the station. Typically, the length of the signal, power and control cables, connecting the SEB to the charged particle detectors, is of the order of 50 m. Figure 3 sketches the connections between the SEB and the HELYCON detectors of the station.

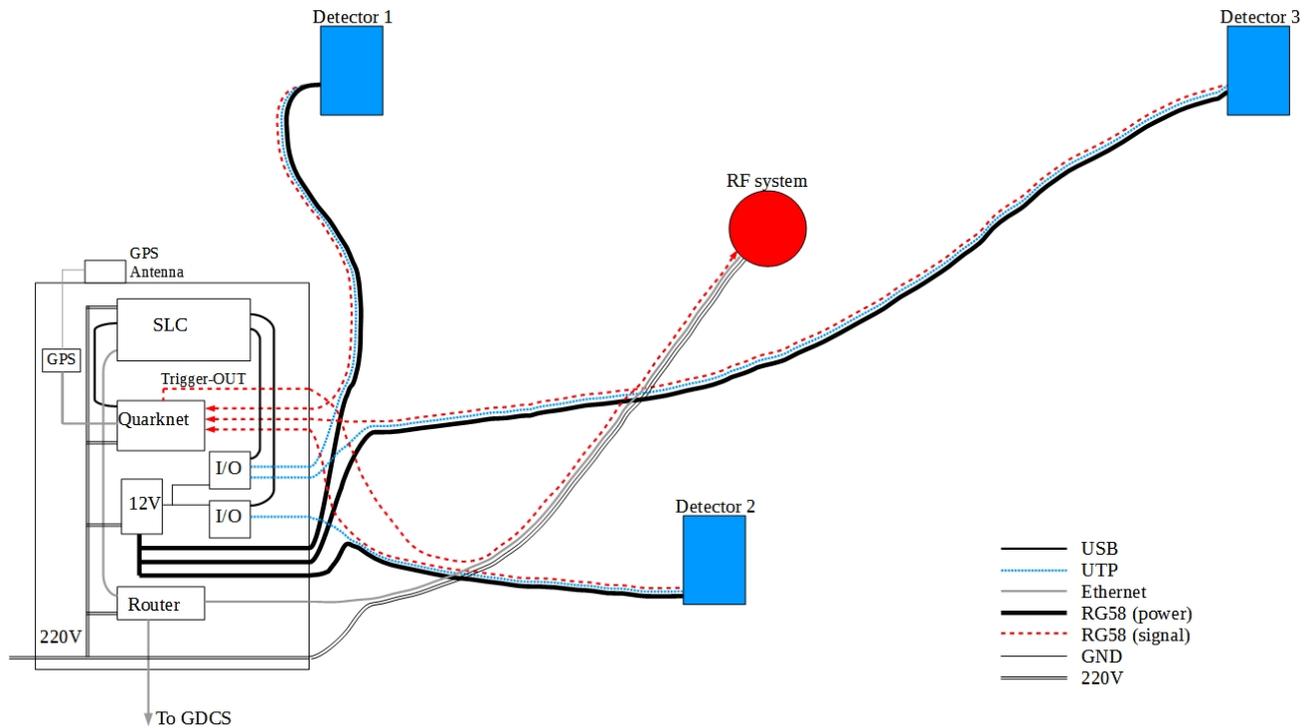

Figure 3: Schematic of the connections between the Station's Electronics Box (SEB) and the station's detectors.

Each of the charged particle HELYCON detectors has been extensively tested and calibrated in the Physics Laboratory of the Hellenic Open University, as it is described in [1, 6, 13]. This included studies of the performance of each PMT (i.e. the characteristics of the pulse height and charge distributions) in detecting single and multiple photons at different values of high voltage, before assembling the detectors. Figure 4 demonstrates charge and peak voltage distribution of

PMT pulses, measured when the PMT is operated at 1.5 kV responding to low intensity light-pulses such that (practically) only single photoelectrons are emitted from the photocathode. It was found that the above distributions were well described by a Gaussian shape with a sigma about 30% of the distribution's mean, which remain practically constant for a wide range of operating voltages. At 1.2 kV, which is the operating voltage of the HELYCON counters, the mean pulse height related to a single photoelectron (pe) is about 2 mV, corresponding to a mean charge of 0.12 pC, varying by less than 10% among the PMTs. At 25° C, the rate of the PMT dark noise does not exceed 40 Hz. At this reference voltage ($V_0 = 1.2$ kV) the PMT gain is typically $G_{V_0} = 7.5 \cdot 10^5$. The dependence of the PMT gain on the operating voltage (V), follows the standard parameterization $G_V = G_{V_0}(V/V_0)^\alpha$, where the parameter α was found to vary by less than 1.3% among the PMTs, around a value equal to 7.05.

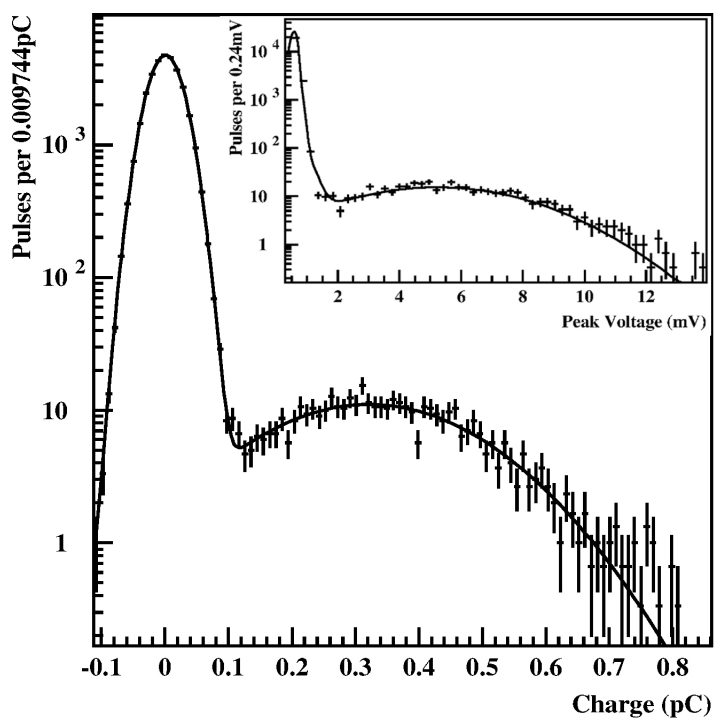

Figure 4: The charge distribution of a HELYCON PMT pulses, when the PMT is operated at 1.5 kV responding to low intensity light-pulses such that (practically) only single photoelectrons are emitted from the photocathode. In the inset plot, the corresponding peak voltage distribution is displayed.

Following the detector assembling, the response (pulse height and charge distributions) of each counter to Minimum Ionizing Particles (MIPs) was measured and their spatial uniformity was tested [1, 3, 6, 13]. The typical response of a HELYCON charged particle detector to a MIP, passing through its geometrical center, corresponds to 21 pe and varies by less than 10% among the different detectors. The detector response to a MIP varies slightly, depending on the position of the intersection point with respect to the PMT position. However, the maximum spatial variation of this response (at the most distant point from the PMT) is less than 15% and it is in agreement with the light attenuation within the optical fibers as it is described by the simulation predictions [3]. In these calibration tests the time delay was measured, relative to the distance of the hit position from the PMT. These delays were found [1, 6] to be consistent with the simulation predictions due to the difference in the corresponding light-path lengths towards the PMT and the refractive index of the wavelength shifting optical fibers.

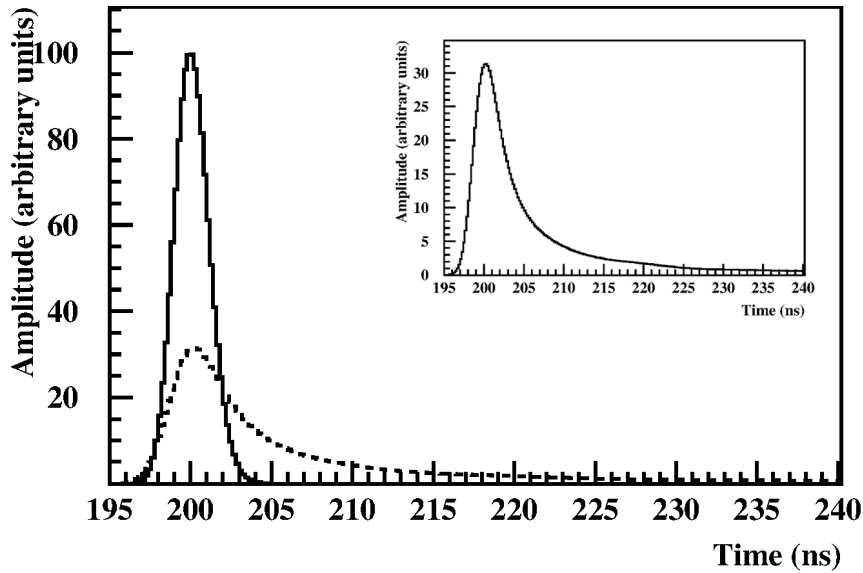

Figure 5: Typical PMT signal at the single pe level, before and after the cable attenuation/deformation. The sharp peak corresponds to the average of 500 individual PMT pulses, digitized in bins of 0.2 ns by a fast (5 GS/s) oscilloscope, with each pulse properly delayed in order to position the peak channel at 200 ns. The inset plot and the dashed curve in the main plot, represent the deformed waveform. Although the integral of the pulse (that is proportional to the charge) is practically conserved, the peak amplitude is reduced to the 30% of the input pulse.

As mentioned above, the detectors' PMT signals are carried to the DAQ electronics by 50 m RG58 cables, which attenuate and deform the waveforms. A narrow test pulse (less than 1.5 ns rise time and 3 ns FWHM) was used in order to quantify this effect by comparing the original waveform with that of the same pulse after going through the whole length of the signal cable. After Fourier transforming the two waveforms, the comparison was made in the frequency domain and the transfer function (amplitude and phase) was determined [14]. The accuracy of this method was tested experimentally by comparing the predicted deformed waveforms to those measured for a variety of test pulses. In order to include the attenuation effects in the simulation description of the detector, it is important to evaluate the waveform of the PMT signal, related to a single pe, after passing through the signal cable. Fig. 5 shows a typical PMT signal at the single pe level, before and after the cable attenuation/deformation. The sharp peak corresponds to the average of 500 individual PMT pulses, digitized in bins of 0.2 ns by a fast oscilloscope, with each pulse properly delayed in order to position the peak channel at 200 ns. The deformed waveform, after the single pe PMT signal passing through 50 m of RG58 cable, was evaluated using the transfer function explained above. This, deformed, shape was used in the simulation (as explained in Section 2.1) to describe the response of the PMT to the scintillation light produced by the particles of a shower front. Although the integral of the single pe pulse (that is proportional to the charge) is practically conserved, the peak amplitude is reduced to the 30% of the input pulse, after passing through the signal cable. However, the peak voltage of PMT's waveforms, corresponding to a collection of many photo-electrons produced by shower particles, is reduced to 60% after passing through the 50 m signal cable. This was tested experimentally by using the signal of HELYCON detector operating in a station, which was duplicated and the two identical pulses were sent to be digitized through a 50 m and 0.5 m RG58 cable, respectively. The distribution of the ratio of the two peak-voltages is in very good agreement with the simulation prediction, which uses the proper single pe pulse shape for each case.

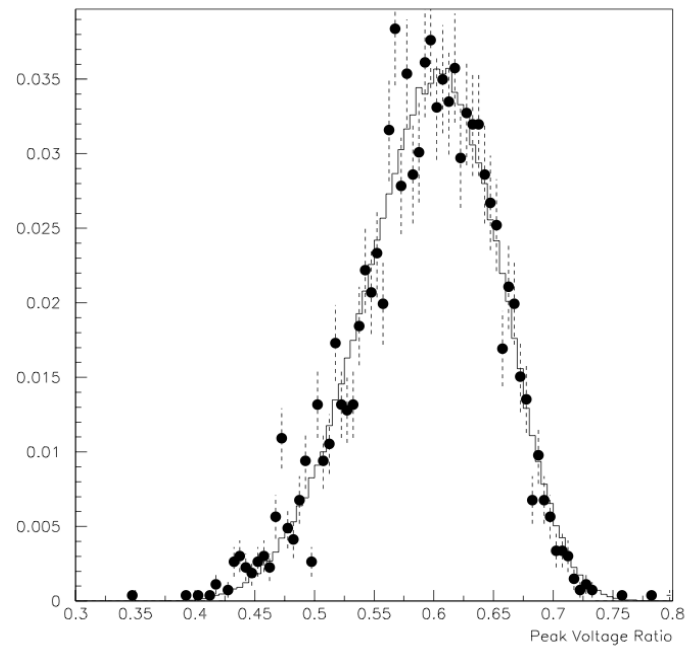

Figure 6: The signal of a HELYCON counter, operating in a station detecting EAS, was duplicated and the two identical pulses were sent to be digitized through a 50 m and 0.5 m RG58 cable, respectively. The solid points represent the distribution of the ratio of the two peak-voltages whilst the histogram corresponds to the simulation prediction.

### 1.2. Control, Digitization and DAQ of the HELYCON station

The operating parameters of the charged particle detectors are managed by online software packages running on the SLC of each HELYCON station. This includes setting of the triggering parameters on the digitization-DAQ card, Quarknet, which is achieved through the DAQ software (described below) and setting the PMTs' high voltages. The latter is controlled by a specifically developed HELYCON-Control software, which provides an interface to the user, via a Graphical User Interface, to select the operating voltage and the corresponding ramp-up time for each of the detectors' PMT. In parallel, it monitors/logs continuously the PMTs' actual voltages and the detectors temperature. In order to set high voltage to a selected detector, the HELYCON-Control software commands the specific output of the station's I/O devices, which is connected to the "control" input of DC-DC converter, that powers the corresponding PMT. In order to monitor, it reads, in regular time intervals selected by the user, the inputs of the station's I/O devices that are connected with the "monitor" output of the DC-DC converters, as well as the inputs that are connected with the output of the temperature sensors in each detector. The HELYCON-Control software displays, in real time, the current values and history plots of the operating voltages and temperatures, providing also alarms and taking emergency actions (e.g. if selected by the user, it shuts down or lowers the voltages) in case that the operating voltages or the detectors temperatures exceed predefined limits.

During a period of almost a year of operating the four HELYCON stations, the recorded monitor information shows that the actual PMT high voltages remain stable. As it is shown in Figure 7 there are small variations, which are strongly temperature correlated but do not exceed 2 V for the whole time period. This corresponds to a variation of the PMT gain of less than 1%.

It is also worth mentioning that, according to the recorded monitor information, the temperature inside the detectors, operating at open field, did not exceeded 37°C during the hottest days of this period, which is the maximum ambient temperature recorded at the same time, due to the insulation of the detectors.

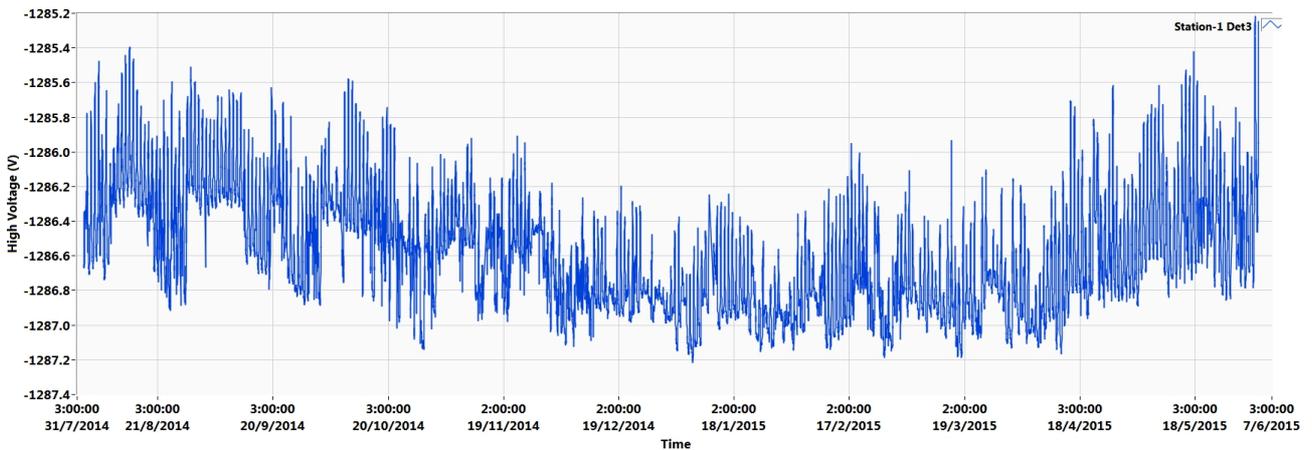

Figure 7: The actual operating voltage values of a PMT as recorded by the HELYCON-Control system for a period of 10 months. The evident daily variations do not exceed, peak to peak, 1.2 V, whilst the maximum seasonal variation is smaller than 2 V. The set high voltage value for this PMT was 1287 V.

The Quarknet board [12], designed and built at Fermilab, is used for event selection, digitization of the signals from the HELYCON counters, data acquisition and time tagging of the events. At each station, the PMT signals of the three charged particle detectors are driven to the three (out of four available channels) inputs of the station's Quarknet board, where after a 10x amplification are compared to a preselected amplitude threshold. Waveforms that exceed this amplitude threshold participate to the trigger formation, which is a majority trigger that requires a certain number of input pulses above threshold to lie inside a time window of preselected width. The timing of the pulses that form the trigger is defined as the earliest time when the pulse waveform crosses the corresponding threshold. The value of the amplitude threshold for each Quarknet channel, the trigger majority level and the duration of the trigger window are selected by the user. When the trigger requirements are fulfilled, the times when the input pulses' waveforms cross the corresponding amplitude thresholds (i.e. the times of rising and falling edge crossings, relative to the trigger) are digitized with an accuracy of 1.25 ns. Each selected event is time-tagged with the absolute arrival time of the latest pulse forming the trigger, using the GPS system of the board, with an accuracy of 40 ns. In parallel, a NIM pulse is generated (hereafter called Quarknet-OUT), which appears at the trigger-output of the board, in a window of 65 ns to 75 ns (equal-probably) after the latest input pulse that participates in the trigger formation. Quarknet-OUT is the pulse which is used to trigger the RF detection system of the station.

The Quarknet board is managed through a specific software, developed at Fermilab in LabVIEW [12], which runs on the SLC. Apart from managing the data transfer to the SLC hard disk, it provides the means to set the operating parameters of the board. These are related to the input channels activation, the amplitude threshold values for each active channel, the trigger majority level, the trigger window width, the maximum time duration of an event as well as to actions such as: start/stop acquisition, monitor the acquisition progress and the event rates, record measurements from the environmental sensors etc. For each triggered event a data stream is saved in a file containing the threshold crossing times of each input waveform, the GPS absolute timing information, as well as environmental and other monitoring information which are measured at preselected time intervals.

Station-4 is run, for calibration and test purposes, with a hybrid digitization and DAQ system using a high sampling rate, 5 GS/s, oscilloscopes (Tektronix 7104) for fully digitizing the detectors signal waveforms in parallel to the Quarknet board, as it is shown in Figure 8. By means of a fast, CAEN Mod. N978 Variable Gain Fast Amplifier [15], analogue amplifier each detector signal is x2 amplified and consequently split by a 50 Ω divider, thus feeding the two parallel systems with identical inputs. The two parallel DAQs run independently of each other and are synchronized by employing the Quarknet-OUT to trigger the fast oscilloscope system. To manage the fast oscilloscope as a multi-channel (up to 4 input channels) waveform digitizer and to transfer and save the collected data, a custom software package [16] has been developed in LabVIEW. Due to the common trigger there is a one to one correspondence between the events collected by the two parallel DAQs, thus offering the possibility to check and calibrate the digitization and other operations of the Quarknet system. Every Quarknet board that is used in a HELYCON station has gone through this evaluation and calibration procedure.

It was found from these tests that there are discrepancies between the set and the actual values of the Quarknet amplitude thresholds, with which the input waveforms are compared. To evaluate the exact relation between the set and actual values, data from EAS were collected with Station-4 using the hybrid digitization system described above, at a certain value of Quarknet set thresholds.

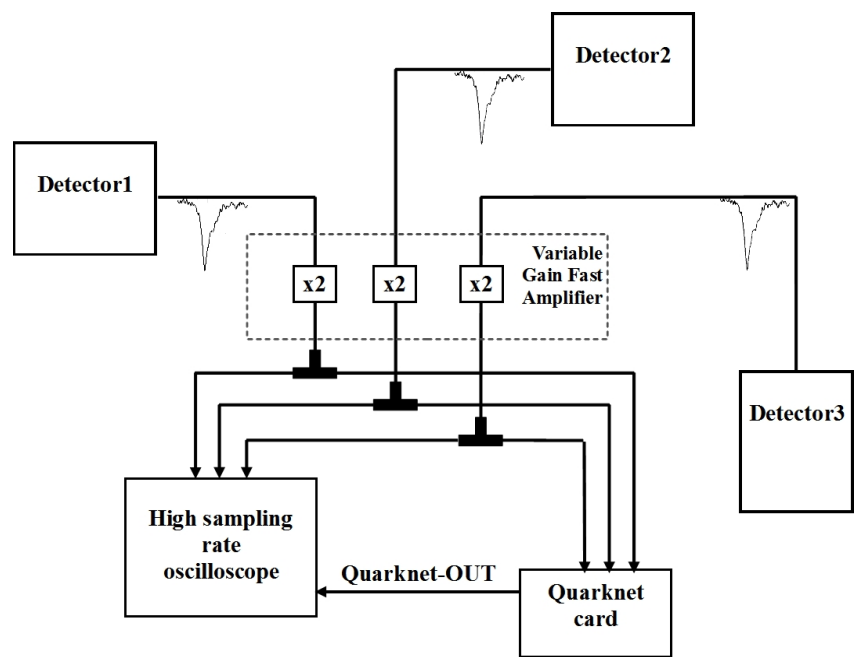

Figure 8: The hybrid digitization and DAQ system used in Station-4

Using the Quarknet data, a Q-ToT (Quarknet Time over Threshold) value was calculated for each pulse, that is the difference between the recorded times when the rising and falling edges of the waveform crosses the preset threshold. The waveforms digitized by the oscilloscope were used for calibration, due to the fact that the actual thresholds were not known accurately. For each Q-ToT, the corresponding waveform, digitized by the oscilloscope in time-bins of 0.2 ns, was analyzed and the width of the waveform (hereafter called Time over Threshold, ToT) at several, accurately known, amplitude levels was evaluated. The actual value of the set threshold of a Quarknet channel was accurately estimated by comparing the Q-ToT values of many input pulses to the ToT values determined from the corresponding, fully digitized waveforms at several amplitude levels.

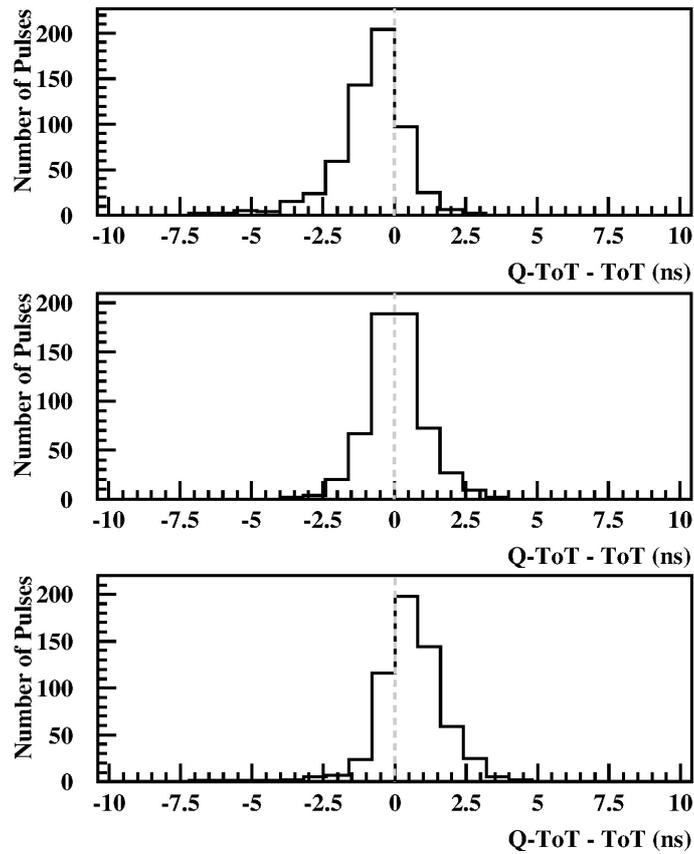

Figure 9: Distributions of the difference between Q-ToT, which is calculated from the Quarknet data and ToT, which is evaluated from the respective fully digitized waveforms. Each distribution corresponds to a different amplitude level, used for the ToT evaluation, i.e. a) 6.1 mV, b) 6.4 mV, c) 6.7 mV. The Quarknet threshold was set to 5 mV.

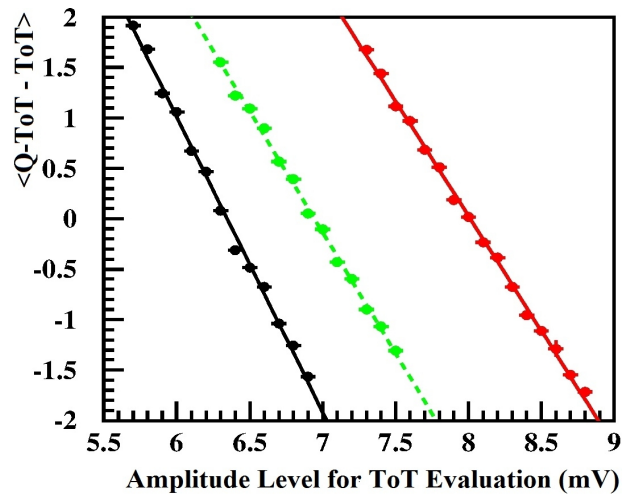

Figure 10. The mean values of the difference between Q-ToT and the respective ToT values as functions of the amplitude levels at which the ToT was evaluated. Each line corresponds to a different Quarknet channel of the same board. At this example the set threshold was 5 mV in all the Quarknet channels. For each channel, the actual value of the set threshold is estimated as the amplitude level where the mean value of the above difference equals zero (card 6811 with thresholds 50-50-50, 41 threshold values with a step of 0.1 mV).

Figure 9 presents distributions of the difference between of the Q-ToT of a Quarknet channel and the respective ToT values, when the ToT has been evaluated at three amplitude levels differing in steps of 0.3 mV. It is apparent that this calibration method can accurately estimate the actual values of the Quarknet threshold. Figure 10 presents, for three Quarknet channels, the mean value of this difference as a function of the amplitude level at which the ToT values were evaluated from the fully digitized waveforms. In this example the thresholds, in all three channels, were set to 5 mV (the HELYCON detectors are operated with low Quarknet thresholds, in the range of 5-10 mV) but the threshold levels at which the ToT values equal the Q-ToT (that is the actual working thresholds) are significantly different. This procedure, of measuring the actual value of the set threshold, was repeated for several values of threshold settings in the Quarknet board, in order to establish calibration curves relating the set to the actual threshold values, for each Quarknet channel. In Figure 11 the distribution of the difference of the relative times of the leading edge of the pulses recorded with the oscilloscope and of the same pulses recorded by the Quarknet card, is presented, indicating a Quarknet leading edge timing resolution of around 0.8 ns.

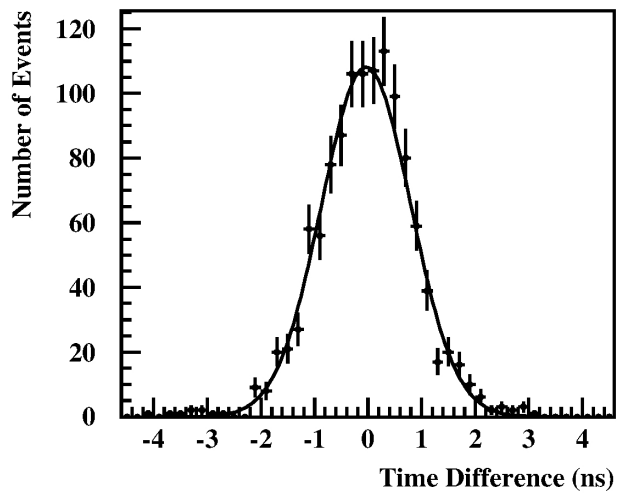

Figure 11: Distribution of the difference of the relative times of the leading edge of the pulses recorded with the oscilloscope and of the same pulses recorded by the Quarknet card. RMS = 0.8 ± 0.03 ns.

## 2. Installation, Simulation and Calibration of HELYCON Counters

In standard data-taking mode, the Quarknet digitization provides information for the relative arrival times (based on the timing of voltage-threshold crossings) and the Q-ToT of the HELYCON counters signals. There is a correspondence between the Q-ToT information and the size (peak voltage and charge) of the detector signals; however, the exact relation depends on the amplitude threshold at which the width of the waveform (i.e. Q-ToT) was measured. Furthermore, the accuracy of the arrival times, due to slewing effects, depends on corrections that are based on both the amplitude of the signal and the threshold level used for timing. A detailed Monte Carlo simulation of the HELYCON counters was developed in order to provide the means for evaluating relation-curves between the above mentioned correlated waveform characteristics, as well as correction curves (e.g. slewing correction) for any set of operation parameters. Calibration data were used to tune the simulation parameters for each individual HELYCON detector and to check the accuracy of simulating the response of each station.

## 2.1 The Calibration Set-Up

The HELYCON stations, before installation, go through a standard calibration and simulation-tuning procedure, operating in a setup as the one depicted in Figure 12. Station-1, installed at its final position, is used as the reference to the calibration setup. Each other Station was operated in a geometrical arrangement where its particle detectors were placed adjacent to the corresponding counters of Station-1. A pair of fast oscilloscopes (7104 Tektronix), recording data with a sampling rate of 5 GS/s, was utilized to digitize the detector signals. The first oscilloscope received inputs from the three counters of Station-1 and recorded data when all the three detector signals exceeded a preselected amplitude threshold. It also provided the external trigger to the second oscilloscope, which digitized the signals of the other Station. The LabVIEW based [16] package has been used for data acquisition. With this setup, full waveforms have been digitized and recorded, corresponding to events when both stations responded to the same EAS. Each digitized waveform comprises 5.000 consecutive measurements of the signal's voltage in time intervals of 0.2 ns. The digitization window, for each counter, was properly adjusted in time, in such a way that the arrival time of the global event trigger corresponded to the 1800$^{th}$ time bin. Two sets, each of about 10.000 fully digitized events, were collected by pairing Station-1 with Station-2 and Station-1 with Station-3, respectively. These data sets were analyzed as described in Section 2.2 and 2.3 in order to fine-tune the simulation parameters.

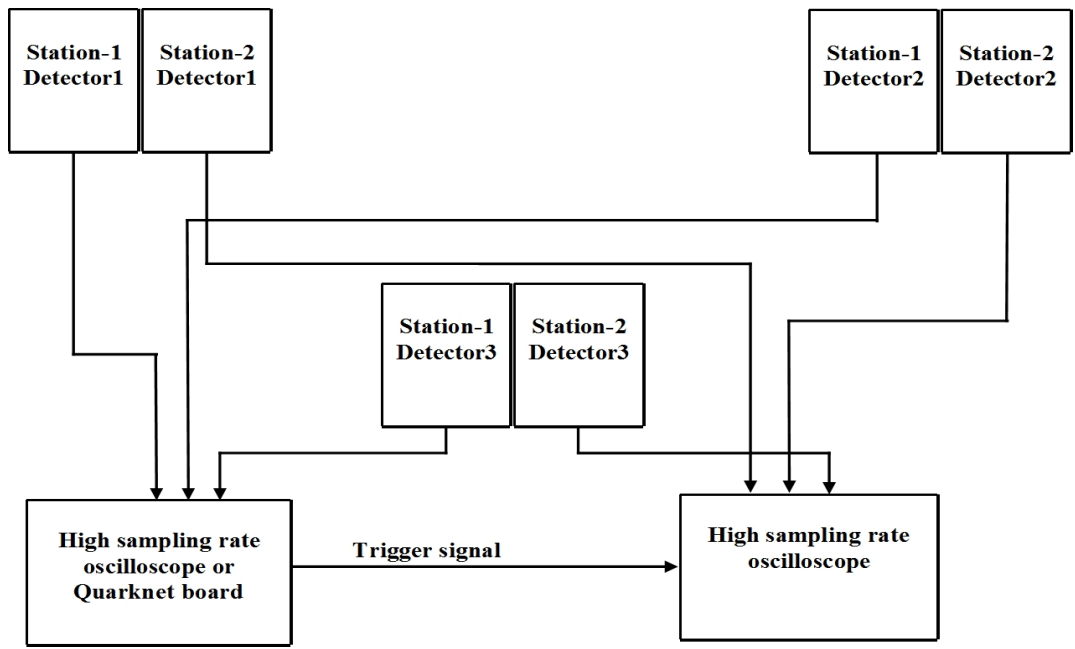

Figure 12: Setup for station calibration and performance evaluation. Two stations are positioned exactly next to each other, recording thus the same EAS events.

Furthermore, in order to crosscheck the timing accuracy of the Quarknet boards, as well as the simulation of their functioning, the reference Station (Station-1) was run with a Quarknet system. The signals of the other Station were fully digitized by the fast oscilloscope, whilst the Quarknet board also provided the external trigger (Quarknet-OUT) to the oscilloscope. All the Quarknet boards, to be used in HELYCON stations, were operated with this setup and the collected data (about 5.000 events per board) have been used for evaluating performances and for comparisons with the simulation predictions.

## 2.2 The HELYCON Simulation

This simulation package is a part of the HOURS (HOU Reconstruction and Simulation) framework [17], which has been developed for Underwater Neutrino Telescopes and for EAS detection arrays. The EAS part comprises simulations of: a) the cascade development using CORSIKA [18], b) the interaction of each particle of the shower's front with the HELYCON counter, c) the production of photoelectrons at the PMT photocathode, c) the detector's waveform formation and d) the functioning of the digitization and triggering electronics.

In generating EAS with CORSIKA, the standard, energy and directional distribution of the primary charged particle was used, whilst the primary's composition follows the distribution described in [19]. The Central European atmospheric model has been chosen whilst the geomagnetic field has been adjusted to the geographical coordinates of the HELYCON site. No thinning was employed for the cascade description. The full EGS, without use of parameterizations, was selected for the development of the electromagnetic part of the shower. The kinetic energy cutoffs of the secondary particles were set to 10 MeV for hadrons and 1 MeV for leptons and gammas. Three sets of CORSIKA simulated showers were used in the following analyses: 1) 200 million cascades in the energy range between $10^{13}$ eV and $5 \cdot 10^{15}$ eV, 2) 50.000 EASs between $5 \cdot 10^{15}$ eV and $10^{18}$ eV and 3) 100 EASs with energies between $10^{18}$ eV and $10^{19}$ eV. Within each set, the energy of the simulated events follows the standard CORSIKA power law distribution. An overall statistical weight was applied to the cascades of each set in order to normalize to the expected number of EASs in a circle of 800 m in radius during a certain time interval, common to all samples.

For each simulated cascade, the shower's particles that exceeded the kinetic energy cutoff and succeeded to reach the ground level were kept. Their energy, the hit coordinates at the ground plane, relative to the impact of the shower axis (hereafter, the shower center), the arrival time to the ground relative to the time of the primary interaction and their direction were saved. Each shower was tried 1400 times by moving the shower center uniformly, inside a circle of 800 m in radius around the center of a HELYCON Station and the response of each counter was simulated. The trials that resulted in detector responses fulfilling trigger criteria were considered as Monte Carlo (MC) events[1]. All the MC events from the same CORSIKA set of cascades share a common weight factor, which is the statistical weight related to the cascade's energy region divided by the number of trials (i.e. 1400).

The particles of the shower front that hit a HELYCON counter were used to simulate the detector's response. At this stage, the simulation employs two types of parameterization tables, which have been produced in a specific, GEANT simulation study[2] of a typical HELYCON detector. The first type of tables contains the distribution of the number of photoelectrons produced, at the photocathode of the PMT, by a single particle of a certain type (e, μ, γ or hadron) and of a certain energy when it hits the detector at a certain point relative to the PMT position. The second type contains the time delay distribution of the produced photoelectrons, relative to the time when a single particle hit the detector, for a certain hit position. At this stage, only time delays related to light's transmission path have been considered, whilst the timing characteristics of the scintillator and of the WLS fibers response were taken into account at a later simulation step.

---

1  Typically, a HELYCON station can observe cascades of the low energy region ($10^{13}$ eV - 5 $10^{15}$ eV) when the shower axis passes by less than 220 m away from its center. This distance for EAS of higher energy ($5 \cdot 10^{15}$ eV - $10^{18}$ eV) is about 420 m whilst for the highest energy considered this distance is about 800 m.
2  As mentioned in Section 2.3 and reported in [1] the performance of this simulation was checked experimentally and it was found to describe well the response of the HELYCON detectors to MIPs, as well as the observed spatial and timing, modest inhomogeneities of the counters.

Figure 13 outlines the stages of simulating the detector response. Starting with the particles that enter the detector and their arrival times (stage a), as determined by CORSIKA, the number of the produced photoelectrons by each particle and their production time is evaluated (stage b), using the relevant distributions from the parameterization tables according to the particle's type, energy and hit position. Consequently, the timing response of the detector (including the time response of the scintillator and the WLS fibers) was taken into account by adding an extra time delay to the production of photoelectrons (stage c), which follows the probability distribution function described by

$$P(t) = R \frac{e^{-t/\tau_1} - e^{-t/\tau_2}}{\tau_1 - \tau_2} + (1-R)\frac{e^{-t/\tau_3}}{\tau_3} \quad (1)$$

where, R and $\tau_1$, $\tau_2$, $\tau_3$ are parameters, which were estimated, for each individual HELYCON detector, by fitting the calibration data, as described in the following Section.

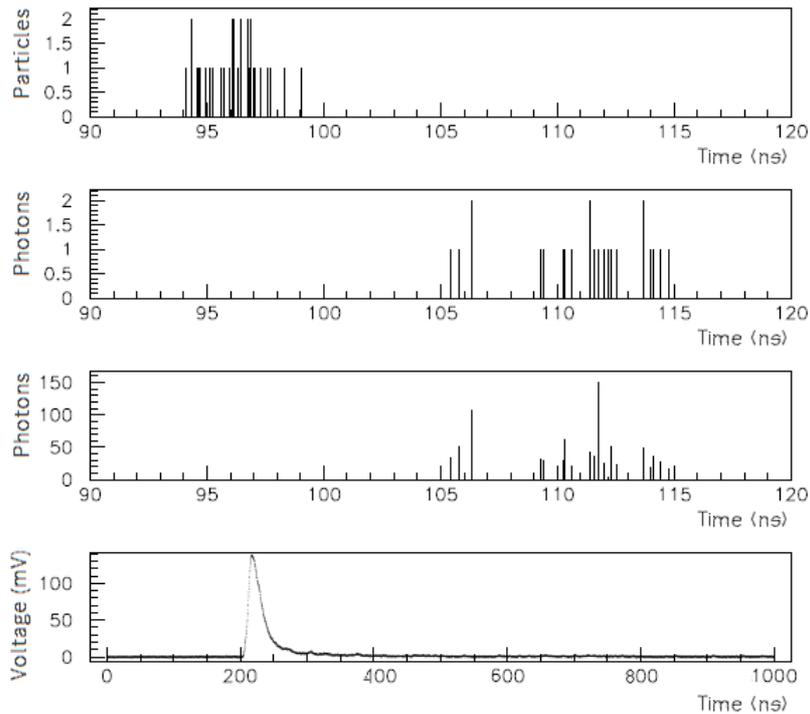

Figure 13: Simulating the detector's response to an EAS particles front. From top to bottom: a) Corsika determines the arrival times of the particles entering the detector, b) Using the parameterization tables described in the text, the total number of produced photoelectrons and their time delay distribution are evaluated, c) Using the detector's time response function (Eq. 1), the photoelectrons are redistributed in time and d) The waveform of the detector's signal as it should be digitized by a fast, 5 GS/s oscilloscope, and a window of 1000 ns and the positioning within the window employed during the calibration. This is evaluated by convolving the time distribution of the photoelectrons with the function that describes a typical PMT pulse corresponding to a single photoelectron, taking into account statistical fluctuations of the single photoelectron's pulse amplitude, the deforming effects of the long signal cables and the oscilloscope electronic noise (see text).

Finally, the waveform of the detector's signal is formed as the sum of pulses, each pulse describing the PMT's response to each of the photoelectrons produced in stage c. These pulses are of the same shape, whilst their amplitude varies, following the charge-distribution that has been determined in laboratory studies [6] of each specific PMT. Deforming effects, due to the long signal cables connecting the detectors to the digitization electronics, have been taken into account by using the deformed pulse shape related to a single photoelectron, determined as described in Section 1.1 and Fig. 5.

Although, the simulation description of each HELYCON detector takes into account the specific PMT characteristics, a common scintillation-light collection efficiency, of 97% has been used for all the counters. However, it was found that this efficiency varies [1] among the different counters and a correction factor (QF), to scale appropriately the amplitude of the simulated waveforms, was evaluated by using the calibration data, as described in the Section 2.3.

Specific routines in the HELYCON Simulation package describe in detail the digitization functions of the Quarknet and the oscilloscope-digitization systems, adding electronic noise when required, as well as emulating the trigger decision electronics and the formation of trigger outputs. When a MC event fulfills the chosen selection criteria and a trigger is formed, the complete digitization information is written to output files, with the same format as the real data. MC event files also contain and information related to the "true" EAS physical parameters as well as other characteristics related to the detectors response.

**2.3 Simulation Tuning and Calibration**

Data, collected by the calibration setup when each of the detector signals of the reference station (Station-1) exceeded 10 mV, were used to tune the simulation parameters. The fully digitized waveforms, in time-bins of 0.2 ns, were processed to evaluate the peak voltage, the corresponding charge of the pulse as well as the pulse width (ToT) at 4.7 mV and 9.7 mV amplitude levels. MC sets of events, produced by simulating the response of the calibration setup to EASs with certain values of simulation parameters, were processed by exactly the same way as the real data.

The simulation parameters (i.e. the light-collection correction factor, $QF$, and the time-response parameters $R$, $\tau_1$, $\tau_2$, $\tau_3$) for each detector of the reference station (Station-1) were estimated by fitting the MC predictions to the charge distribution and the average waveform of the detectors signals. The events used in this fit were selected from the data and MC sets by requiring each of the detectors signal to exceed a threshold of 12.7 mV. However, in the following analysis only detector signals with peak-voltage less than 180 mV were used, in order to avoid overflows caused when the peak voltage exceeds the selected oscilloscope scale. For each individual detector, the charge distribution of the signals was established as a histogram and the mean functional form, V(t), of the detector's signal was evaluated as a function of the related charge. This mean detector's signal was estimated by averaging[3] those of the fully digitized waveforms that correspond to charges lying within a certain charge interval (for a demonstration see Fig. 17). To cover the entire charge range, up to charges corresponding to signals produced by 90 MIPs, a total number of 20 charge intervals were used.

The estimation of the simulation parameters followed an iterative procedure of two steps. In a first step, only values for the $QF$ parameters were estimated, whilst some initial values of the time

---

3  The waveforms were synchronized before averaging by shifting them in time in such a way that their leading edges cross a threshold level of 2 mV at the same time.

response parameters were assumed, by maximizing the following, binned, extended likelihood function:

$$L_{QF} = \prod_{\substack{i=1,N_b \\ k=1,3}} \frac{[\mu_{ki}(QF_1,QF_2,QF_3;S_1,S_2,S_3)]^{n_{ki}} \cdot e^{-\mu_{ki}(QF_1,QF_2,QF_3;S_1,S_2,S_3)}}{n_{ki}!} \quad (2)$$

where the charge of each counter was binned in $N_b$ bins, $k$ denotes the detector, $QF_k(k=1,2,3)$ is the light-collection correction factor for the kth detector, $n_{ki}$ are the number of observed pulses of the $k_{th}$ detector with corresponding charges in the $i_{th}$ bin $(i=1,2,3,\ldots N_b)$, $\mu_{ki}(QF_1,QF_2,QF_3;S_1,S_2,S_3)$ is the MC expected occupation of the bin $i$, when the light-correction factors are $QF_1,QF_2,QF_3$ the light-collection and the event selection conditions are the same as the real data. The MC prediction of the charge distribution of a counter $(\mu_{ki}, i=1,2,3,\ldots N_b)$ depends on the assumed $QF$ values of all the three detectors, due to the fact that the $QF$ factors influence the amplitude of the simulated pulses that participate in the event selection trigger. For the same reason the expected charge distribution of each counter depends on the three sets of time response parameters $S_k \equiv \{R_k, \tau_{1k}, \tau_{2k}, \tau_{3k}\}, k=1,2,3$, which determine the shape of the PMT pulses.

In a second step, the above estimated $QF$ values were used to produce MC samples of events, each sample using different sets of values for the time response parameters. At this stage the 20 mean-signal forms of an individual detector, each related to a different pulse-charge region, were compared with the corresponding MC predictions in order to estimate the values of the time response parameters that describe best the detector's waveform for the entire charge range. It was found that a very good description of all detectors signals could be achieved by letting the parameters $R$ and $\tau_1$ to vary as functions of the waveform's charge, $Q$, as:

$$\begin{aligned} R_k &= a_{Rk} + b_{Rk} \cdot Q \\ \tau_{1k} &= a_{1k} + b_{1k} \cdot e^{-c_{1k}Q} \end{aligned} \quad k=1,2,3 \quad (3)$$

whilst $\tau_{2k}$ and $\tau_{3k}$ remain independent of the charge ($k=1,2,3$). In using the above parameterization in the simulation, through the time smearing described by Eq. (2), it was assumed that $Q$ equals the sum of the charges of the single-photoelectron pulses used in the formation of the waveform. The time response parameters, for the kth detector, were estimated by minimizing the chi-squared defined in Eq. (4) as:

$$\chi_k^2 = \sum_{\substack{i=1,N \\ j=1,20}} \frac{\left[V_{ijk} - U_{ijk}\left(\overbrace{a_{Rk}+b_{Rk} \cdot Q_j}^{R_k}, \overbrace{a_{1k}+b_{1k} \cdot e^{-c_{1k}Q_j}}^{\tau_{1k}}, \tau_{2k}, \tau_{3k}\right)\right]}{\sigma_{ik}^2} \quad (4)$$

where the index $i(1,2,3,\ldots,N_s)$ denotes the time bin of the waveform, the index $j(1,2,3,\ldots,20)$ denotes the charge region around a central value $Q_j$; $V_{ijk}$ is the height, at the $i_{th}$ time bin, of the mean-waveform of data signals with charge at the $j_{th}$ charge region; $U_{ijk}$ is the MC prediction for the height of the $i_{th}$ time bin at the $j_{th}$ charge region, corresponding to following set of time response parameters $S_k = \left\{ \overbrace{a_{Rk}, b_{Rk}}^{R_k}, \overbrace{a_{1k}, c_{1k}}^{\tau_{1k}}, \tau_{2k}, \tau_{3k} \right\}$

Using the estimated values of the time response parameters for each counter and their established dependence on the charge of the signal, the two step procedure was repeated until the estimations of the light-collection correction factors and the time response parameters to converge, practically after the third iteration. The simulation parameters of the second station of Fig. 13 were estimated by the same procedure as in Station-1. The event-selection criteria were applied, as before, on the reference station and detector pulses of the second station with peak voltage between 12.7 mV and 180 mV were used in the fits. The estimated QF values vary among the HELYCON detectors between 0.9 and 1.1, whilst the estimation error is of the order of 0.03. The estimated values of the time response parameters found to be similar among the detectors with typical values as:

$$R = \left[ 0.75 + 6.5 \cdot 10^{-4} \cdot N_{MIP} \right] ns$$

$$\tau_1 = \left[ 10.7 - 5.1 \cdot e^{-0.15 \cdot N_{MIP}} \right] ns$$

$$\tau_2 = 9.3 \, ns$$

$$\tau_3 = 670 \, ns$$

where $N_{MIP}$ is the charge of the PMT pulse in units of the charge of a typical detector response to a minimum ionizing particle.

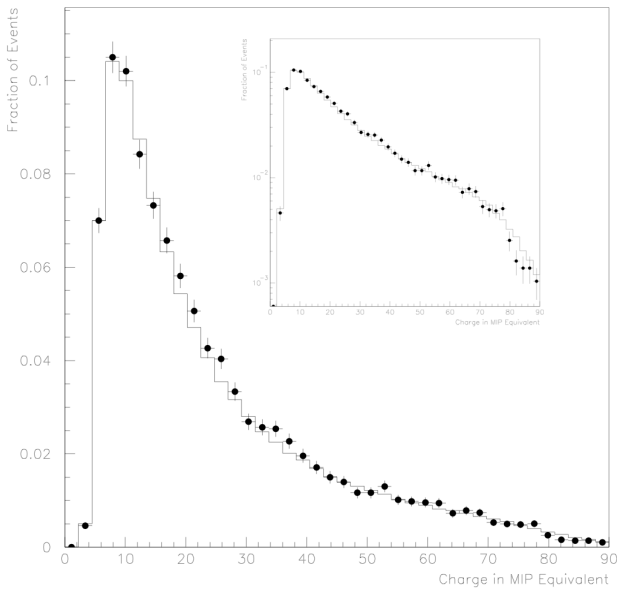
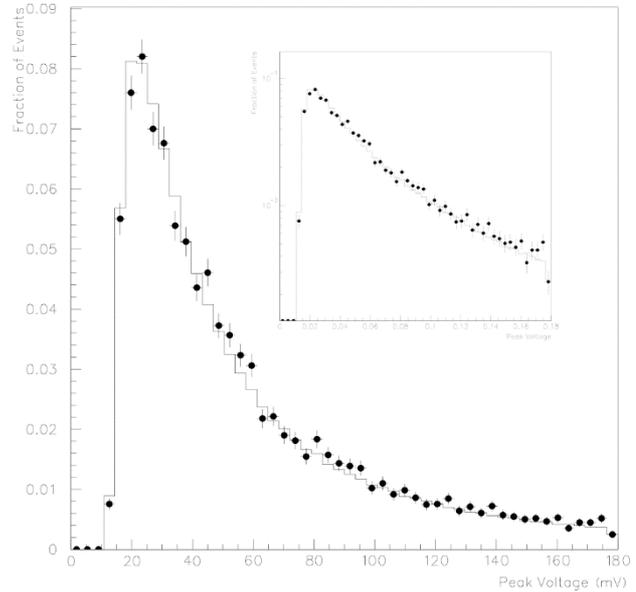

Figure 14: The charge distribution of a HELYCON detector (points) operating in Station-1 of the calibration setup, in comparison with the MC prediction (histogram) after tuning the simulation parameters. Both linear and logarithmic (inset) scales are shown.

Figure 15: The peak-voltage distribution of a HELYCON detector (points) operating in Station-1 of the calibration setup, in comparison with the MC prediction (histogram) after tuning the simulation parameters. Both linear and logarithmic (inset) scales are shown.

Demonstrations of the MC performance, after tuning, to describe the detectors response are shown in Figures 14-17. Specifically, Fig. 14 and Fig. 15 present the charge and peak-voltage distributions of a detector's signals, operating in Station-1, which are in very good agreement with the MC predictions. Fig. 16 demonstrates the agreement between the calibration data and the MC prediction with respect to the ToT distribution of signals, collected by a detector operating in the second station of the calibration, whilst Fig. 17 presents the averaged signal waveforms of a detector of the Station-3 in comparison with the MC prediction. The success of the MC to express the peak-voltage and the ToT distributions of the HELYCON detectors, whilst these waveform characteristics have not been used in trimming the predictions, indicates that the tuned MC describes suitably the detectors response.

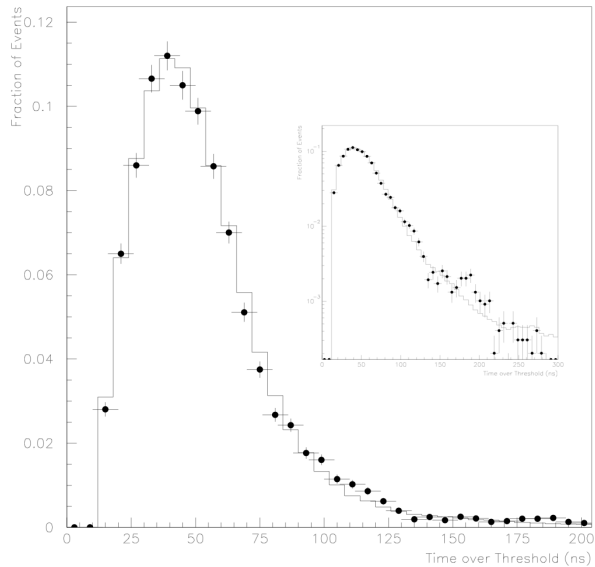

Figure 16: The distribution of ToT at 9.7 mV, of a HELYCON detector (points) when operates in the second station of the calibration setup, in comparison with the MC prediction (histogram) after tuning the simulation parameters.

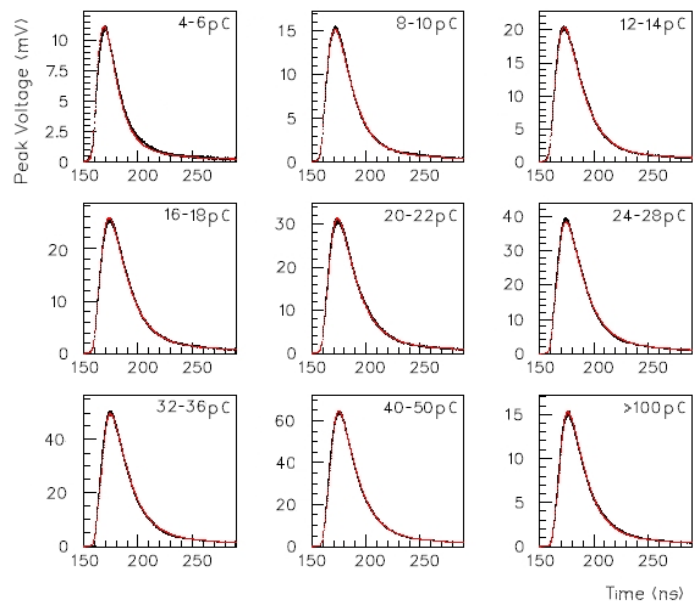

Figure 17: Comparison of mean signal waveforms, evaluated by averaging digitized waveforms of a HELYCON detector operating in the Station-3 (points - red), with the MC predictions (line - black) after tuning the simulation parameters, for signals with charge in the areas noted on each plot.

In order to evaluate the performance of the tuned MC to describe the correlations between the relevant waveform characteristics (i.e. charge, peak-voltage and ToT), the detectors signals have been categorized in bins of their ToT values, measured at a certain (4.7 mV or 9.7 mV) threshold level. In each of the above ToT-classes, the mean values of the charge and the peak–voltage were evaluated, establishing thus the relationships that express the correlations of the signal's charge and peak-voltage with the corresponding ToT. As it is demonstrated in Fig. 18 and Fig. 19, the above relationships observed in the calibration data are very well described by the tuned MC predictions. It was found that this agreement holds for all the calibrated detectors. Additionally these relationships were found to be almost the same among the different HELYCON counters, when run with the same gain.

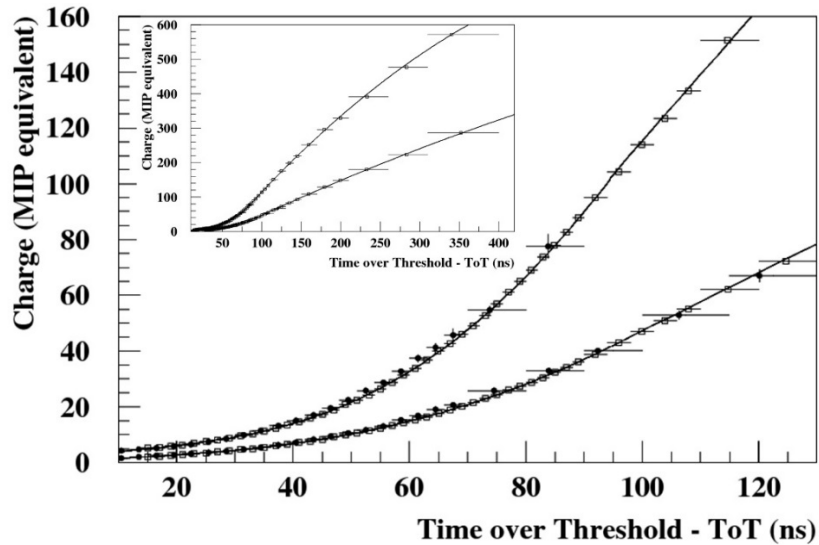

Figure 18: The relation between the charge of a HELYCON detector' signals with the corresponding ToT, when ToT is measured at: a) 4.7 mV and b) 9.7 mV threshold levels. The solid points and the open circles represent calibration data and MC predictions, respectively. The lines are graphic representations of functional fits to the MC points. The inset plot contains the MC predictions for a wide range of ToT values.

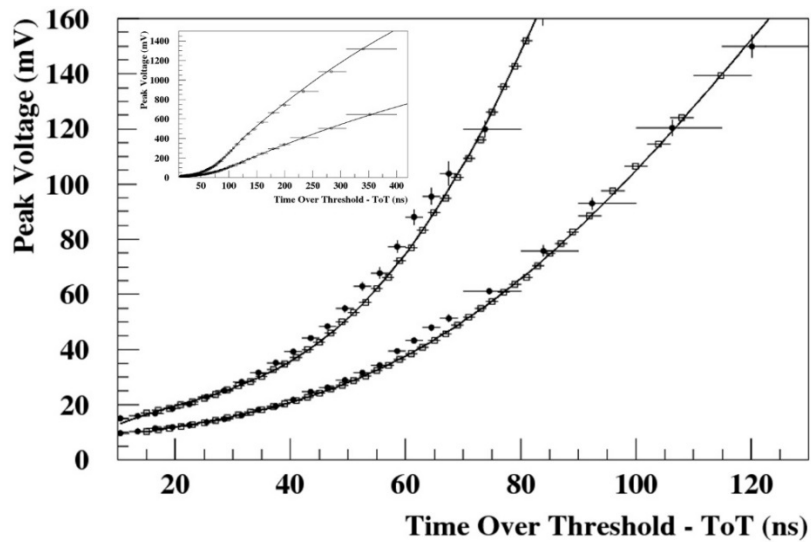

Figure 19: The relation between the peak-voltage of a HELYCON detector' signals with the corresponding ToT, when ToT is measured at: a) 4.7 mV and b) 9.7 mV threshold levels. The solid points and the open circles represent calibration data and MC predictions, respectively. The lines are graphic representations of functional fits to the MC points. The inset plot contains the MC predictions for a wide range of ToT values.

Although the calibration data do not provide information[4] for very large ToT values, the observed

---
4  Which is mainly due to the oscilloscope overflow for large pulses but also because very large detector signals are infrequently agreement between calibration data and MC predictions justifies the use of the simulation to establish functional relations that describe the correlation between the mean values of the charge and the peak-voltage with ToT, for each calibrated detector, up to larger than the observed ToT values, as shown graphically in the inset plots of Figures 18 and 19. These parameterizations, especially the charge vs ToT relation, are used in the data analysis [20] to estimate the magnitude of the detectors signals, when the HELYCON stations are operating with the Quarknet electronics and only the ToT measurement is available.

Such estimation, based on the above parameterizations (hereafter called "charge estimation-curves"), suffers from statistical errors that also were quantified using the calibration data and the same procedure was repeated using simulated events in order to establish the MC predictions. Specifically, the measured ToT value of each digitized detector's signal was used to estimate the charge of the waveform using the corresponding estimation-curve, whilst the "true" charge of the same signal was calculated by integrating the fully digitized waveform, thus establishing the deviation of the estimated charge from its true value. The detector signals were then categorized in bins of the estimated charge. The mean value and the RMS of the above deviations, evaluated in each bin of the estimated-charge, were used to check the consistency and the accuracy of the charge-estimation using ToT. The mean values of the deviations found to be consistent with zero in all the estimated-charge bins, indicating a practically unbiased estimation, whilst their RMSs represent measures of the estimation error. The RMSs, expressed as fractions of the estimated charge, are presented in comparison with the MC predictions in Fig. 20, where the lines represent functional fits to the MC error predictions. Despite the general agreement between data and MC predictions, the estimation errors evaluated from the calibration data, especially at larger pulses, are slightly but systematically higher than the simulation predictions. These small discrepancies are due to the fact that the ToT measurements suffer from electronic noise fluctuations, which are not completely embedded in the simulation. Consequently, the charge estimation based on ToT measurements of the calibration data are distributed around the true charge value but with larger variance than predicted by the MC. However, these disagreements are much smaller than the MC predicted RMSs, justifying the use of the MC predictions to quantify the expected estimation error when estimating the charge of the detector signals using ToT measurements. The solid lines in Fig. 20 represent fits of the MC points, which provide the parameterization of the expected charge-estimation error as a function of the estimated charge. Charge estimation curves and the related charge-estimation error parameterizations were thus established for every one of the HELYCON detectors, corresponding to the working high voltage of each detector, at 4.7 mV and 9.7 mV threshold levels.

It was found that, in all the calibrated detectors, the charge estimation based on ToT measurements at 9.7 mV results to slightly better accuracy, for signals corresponding to less than 40 MIPs, than when ToT is evaluated at the lower threshold of 4.7 mV, as it is also shown in Fig. 20. Such a behavior is expected since the main error-source in measuring ToT is due to waveform fluctuations, which are more evident at low voltage levels, especially at small and medium size signal pulses. However, for larger pulses the estimated charge is slightly more precise when the ToT is evaluated at 4.7 mV threshold level. Although waveform fluctuations of large pulses affect practically the same the ToT measurements, taken at 4.7 or 9.7 mV thresholds, practically the same observed in the collected calibration data.

way, the curve describing the charge vs ToT relationship is steeper when ToT is measured at 9.7 mV than the corresponding curve at 4.7 mV, as it is shown in Fig. 18. Consequently, even when the ToT measurement error is the same, the higher derivative results in larger charge-estimation error at 9.7 mV than in estimations based on ToT measurements at 4.7 mV threshold level.

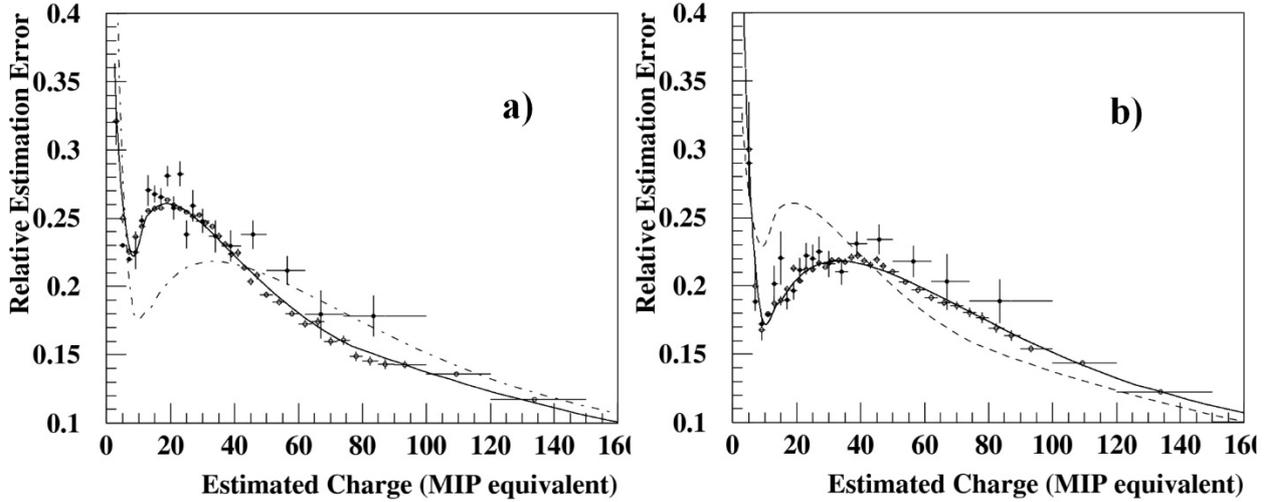

Figure 20: The error of charge estimation, expressed as fraction of the estimated charge (relative estimation error), versus the estimated charge; a) when the charge estimation is based on ToT measurements at 4.7 mV threshold level and b) when ToT is measured at 9.7 mV threshold level. The solid points represent the estimation errors evaluated using calibration data of a detector of Station-1, whilst the open circles denote the corresponding errors evaluated using simulated events. The charge-estimation error is defined as the RMS of the deviation of the estimated charge from the corresponding true charge, when the estimated charge value lies in a certain region symbolized by a horizontal bar in the plots. The solid line in a) and the dashed line in b) represent the fit to the MC points when ToT is defined at 4.7 mV, whilst by a solid line in b) and a dashed line in a) graphically represents the fit to the MC points when ToT is measured on 9.7 mV threshold.

**2.4 Timing Studies and Calibration**

The calibration set up, shown in Fig. 12, was used to evaluate the resolution in measuring the arrival time of the shower's front in each HELYCON counter. This study uses data collected as described in Section 2.1 also fulfilling an extra requirement that the detectors' signals, of both Stations, exceed in amplitude a threshold of 10 mV. The arrival time of a detector's signal is defined as the time when the leading edge of the fully digitized waveform crosses a predefined threshold level (hereafter timing threshold level), in the same way that the Quarknet electronics time the detectors pulses. It is known that this definition introduces systematic errors (slewing) in measuring arrival times, which also depend on the pulse amplitude. This analysis aims: a) to quantify the slewing systematics in a way that can be used to correct the timing on a pulse by pulse basis, b) to measure the timing resolution of the detectors as functions of the signal amplitude and c) to evaluate the detector simulation performance in describing the above timing characteristics.

It is assumed that, due to proximity of the adjacent counters, the particle-front of the extensive shower reaches their surfaces simultaneously, resulting to synchronous detector signals. Let $\delta t$ (=$t_2$-$t_1$) be the difference in arrival times of the two signals from a pair of adjacent detectors, belonging to the Station-2 and the reference Station-1 respectively, when responding to the same EAS. Assuming synchronous responses, $\delta t$ should be distributed around zero independently of the elapsed time between $t_1$ and the formation of the event selection trigger $(t_{trg})^5$. However, the above assumption is only approximately valid. Due to the non-negligible size of the HELYCON counters (1x1 m$^2$) there is a time lug, of the order of ns, to the time response of adjacent counters. This time shift depends on the direction of the particles shower relative to the geometrical orientation of the corresponding detectors. This is demonstrated at the left column of Fig. 21, for all the three detector-pairs of the calibration set-up. The arrival time difference (δt), defined as above for a pair of counters, is plotted versus $t_1-t_{trg}$, i.e. the time delay of the detector signal that belongs to the reference Station-1 ($t_1$) with respect to the arrival time of the selection trigger ($t_{trg}$). For these plots, only pulses higher than 40 mV have been used (in order to avoid large slewing systematic errors) and the relevant times have been corrected for constant delays due to different cable lengths.

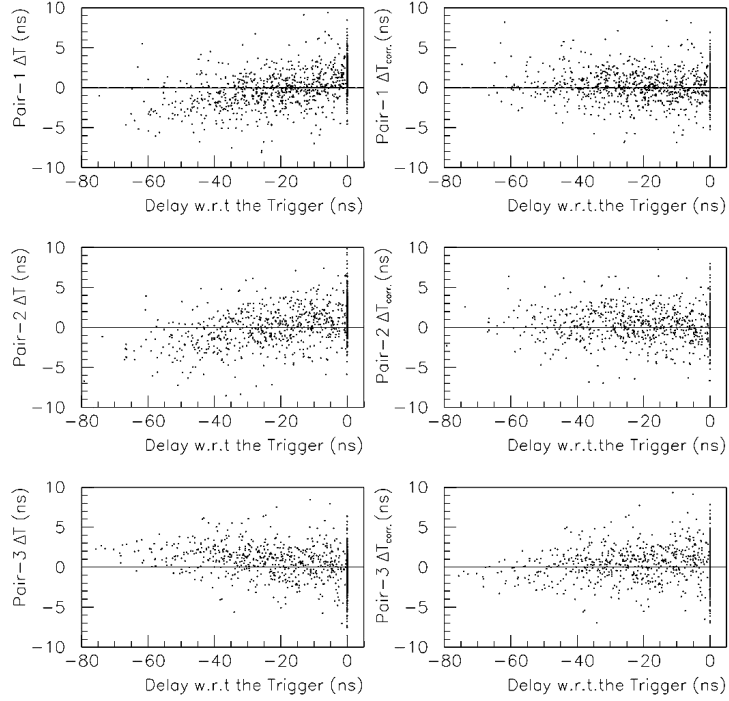

Figure 21: The difference in time response (δt) of two adjacent detectors, belonging to the Station-2 and the reference Station-1 respectively, when both detect the same EAS, is plotted versus $t_1-t_{trg}$, which is the time delay between the signal of the detector that belongs to the reference Station-1 and the event selection trigger. The plots at the left column correspond to the three pairs of adjacent counter of the calibration set-up before any corrections applied to the time differences whilst the right column plots are after the application of empirical corrections evaluated by the simulation and explained in the text.

The mean δts for all the detector pairs seem to vary, depending almost linearly on the corresponding delay ($t_1-t_{trg}$) w.r.t. the trigger. Applying the same timing analysis on simulated events, exactly the same trends as in the calibration data have been observed. Furthermore, MC studies show that the slope of these linear trends depend on the geometrical arrangement of the pair of adjacent counters relative to the other counters of the calibration set-up. High statistics samples of MC events have been used to parameterize linearly this deviation of the mean δt from zero, as a function of ($t_1-t_{trg}$), the delay relative to the trigger. These timing corrections are specific for each calibration set-up and each pair of adjacent counters (i.e. the geometrical position within the calibration set-up). The effect of applying these corrections to the calibration data is demonstrated at the right column's plots of Fig. 21. After correcting, the average differences in time response (δt) of adjacent counters stays constant, around zero, independently of the delays relative to the trigger. Throughout the following timing studies, these corrections (hereafter referenced as "geometrical timing corrections") were always applied before any comparison between the time responses of adjacent counters.

---

5   Notice that the event selection trigger is formed by the triple coincidence of the detector signals of the reference Station-1, when each of them exceeds 10 mV in amplitude, with the latest pulse defining the timing of the trigger. Consequently, the time difference $t_1-t_{trg}$ depends on the direction of the shower and the geometrical positions of the counters.

The systematic (slewing) and the statistical (resolution) errors of measuring the arrival time of a detector's signal depend on the signal's amplitude [1]. Due to the fact that the only available experimental information related to the signal-amplitude is the signal's ToT at a certain threshold, the timing resolution and slewing have been evaluated as functions of ToT. The calibration data and the MC events have been analyzed in exactly the same way, in order to evaluate the timing errors, at 4.7 mV and 9.7 mV timing-threshold levels, as functions of the related ToTs.

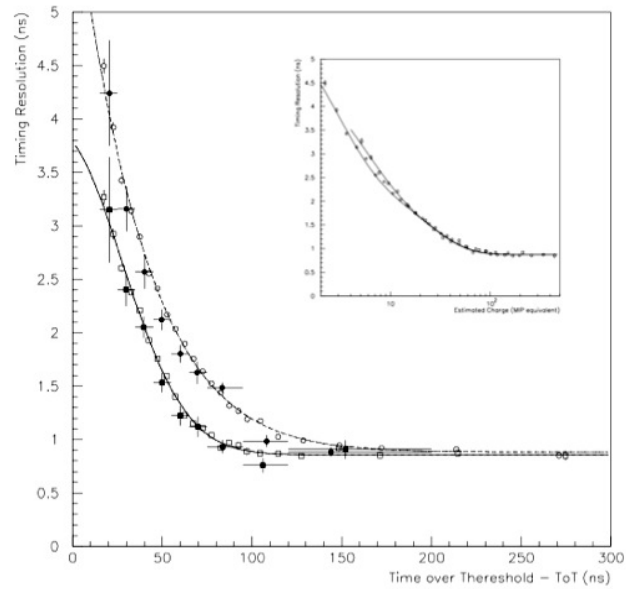

Figure 22: The timing resolution, evaluated by comparing the time response of a pair of adjacent counter as described in the text, as a function of the ToT of the signals. The circles and the squares represent the resolution of timing measurements performed at 4.7 mV and 9.7 mV timing-threshold levels, respectively. Solid symbols (circles and squares) and open symbols correspond to results obtained by analyzing calibration data and MC events respectively. The inset plot presents the MC results on the detectors timing resolution as functions of the signals charge, where the corresponding charge-estimation curves were used to transform the ToT values to charge.

The timing resolution was estimated in bins of ToT by exploiting the spread of δt distributions, where both the signals of the adjacent detectors have similar ToT values that belong to the same bin. By using narrow ToT-bins, the arrival-time measurements of both signals ($t_1$ and $t_2$) suffer practically of the same slewing, resulting to the cancelation of systematic effects when forming the time difference $\delta t = t_2 - t_1$. Indeed, as found in both the calibration data and the MC events, these δt-distributions in bins of ToT are well approximated with Gaussians centered at zero, for all the used bins of ToT. Assuming the same timing resolution for each of the adjacent counters, the timing resolution was evaluated in bins of ToT $\left(ToT \in [T_H - d, T_H + d]\right)$ as $\sigma(T_H) = \sigma_{pair}/\sqrt{2}$, where $T_H$ is the signal ToT mean value at threshold H, d is the half of bin width and the $\sigma_{pair}$ parameter was estimated by a Gaussian fit of the corresponding δt-distribution. The dependence of the timing resolution of a typical HELYCON detector on the signal's amplitude (ToT), evaluated using the calibration data, is shown in Fig. 22 in comparison with the corresponding MC predictions.

As shown in Fig. 22, the simulated response of the HELYCON counters describes very well the measured timing resolution of the detectors. Similar dependences and the same good agreement between the MC predictions and the timing resolutions evaluated from the calibration data was found to hold for all the calibrated HELYCON detectors. The apparent different dependences of the timing resolution on the ToT when 4.7 mV or 9.7 mV threshold were used, is due to the fact that the same ToT value corresponds to different signal amplitudes at different threshold levels. This is demonstrated on the inset plot of Fig. 22, where the timing resolution is plotted as a function of the estimated charge of the signal, by using the ToT measurement to estimate the signal's charge, as it is described in the previous Section. Practically, for a certain signal amplitude (charge) the same timing resolution can be achieved using either one of the above thresholds.

In order to estimate the slewing systematics, the probability distribution function (pdf) $R_H(t_m; t_R, \sigma(T_H), \Delta(T_H))$ that expresses the resolution of a measurement of the signal's arrival time ($t_m$) at a certain threshold level (H), when the «true» value of the arrival time is $t_R$ and: i) the amplitude of the signal corresponds to a certain ToT value ($T_H$), ii) the timing resolution is known as a function of the corresponding ToT ($\sigma(T_H)$), iii) the systematic shift due to the slewing is also a function of the ToT value ($\Delta(T_H)$), was expressed as:

$$R_H(t_m; t_R, \Delta(T_H)) = \frac{1}{\sqrt{2\pi}\sigma(T_H)} \cdot e^{\frac{-(t_m - \Delta(T_H) - t_R)^2}{2\cdot\sigma^2(T_H)}} \quad (5)$$

where the function $\sigma(T_H)$ has been evaluated as described in the previous paragraph.

Assuming that independent time measurements, $t_m^1$ and $t_m^2$, of the arrival of a single EAS by a pair of adjacent counters correspond to a common «true» arrival time $t_R$, the time difference $\delta t = t_m^1 - t_m^2$ distributes according to the following pdf:

$$R_H(\delta t; \Delta(T_H^1), \Delta(T_H^2)) = \frac{1}{\sqrt{2\pi(\sigma^2(T_H^1) + \sigma^2(T_H^2))}} \cdot \exp\left(\frac{-(\delta t - \Delta(T_H^1) + \Delta(T_H^2))^2}{2\cdot(\sigma^2(T_H^1) + \sigma^2(T_H^2))}\right) \quad (6)$$

where $T_H^1$ and $T_H^2$ are the ToT values of the two signals at threshold $H$ and it is also assumed that the two counters follow the same statistical and systematical timing error dependence (i.e. $\sigma(T_H)$ and $\Delta(T_H)$) on the amplitude of the signal.

In the following, $\Delta(T_H)$ was parameterized as $\Delta(T_H) = p_H \cdot e^{-q_H \cdot T_H}$ and the parameters $p_H$ and $q_H$ were estimated by an unbinned likelihood fit using the whole of calibration data, i.e. by minimizing, with respect to $p$ and $q$, the negative logarithm of the following likelihood function

$$L(p,q) = \prod_{i=1}^{N} \frac{1}{\sqrt{2\pi(\sigma^2(T_{H_i}^1) + \sigma^2(T_{H_i}^2))}} \cdot \exp\left(\frac{-(\delta t - \Delta(T_{H_i}^1; p_H, q_H) + \Delta(T_{H_i}^2; p_H, q_H))^2}{2\cdot(\sigma^2(T_{H_i}^1) + \sigma^2(T_{H_i}^2))}\right) \quad (7)$$

$$\Delta(T_H; p_H, q_H) = p_H \cdot \exp(-q_H \cdot T_H)$$

where $N$ is the total number of the calibration events used in the fit. The $p_H$ and $q_H$ parameters were evaluated for all the calibrated pairs of HELYCON counters at both reference thresholds (4.7 and 9.7 mV) resulting to similar results for all detectors. Indicative values of the parameters (at 9.7 mV) are $p_H \sim 16.05$ ns and $q_H \sim 0.034$ ns$^{-1}$ which correspond to a slewing of the order of 10 ns for the smallest recorded pulses which drops to less than 1 ns for pulses greater than 110 mV. The MC predictions for the slewing systematics (by applying exactly the same analysis to the simulated events) were found in a very good agreement with the above results. The estimated function $\Delta(T_H; p_H, q_H)$ is used to correct the timing measurements of each HELYCON detector as a function of the measured ToT, i.e. $t_{corrected} = (t_m - \Delta(T_H))$ .

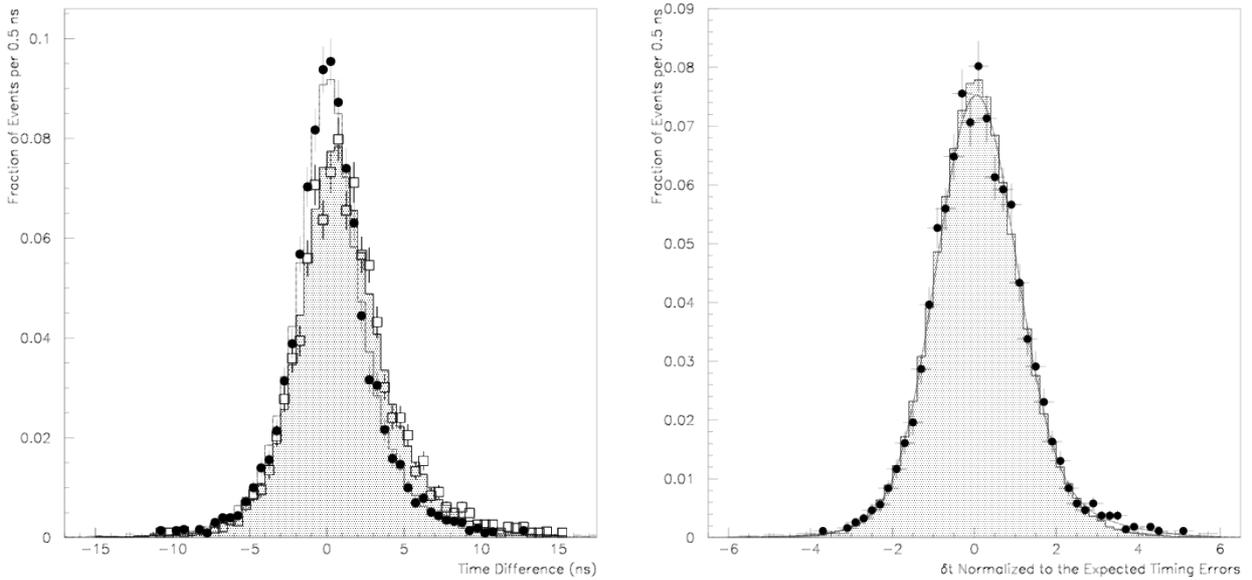

Figure 23: Left: The difference in arrival times of signals from a pair of adjacent detectors, before applying corrections (open squares and shadowed histogram) and after the application of corrections (solid circles and not-shadowed histogram). Points represent calibration data while histograms correspond to MC predictions. The timing has been performed at DAQ 9.7 mV threshold level. Right: The respective pull distributions for data and MC predictions. The solid line represents a Gaussian fit to the calibration data resulting to a mean and sigma parameter values consistent with zero and one respectively.

Fig. 23 demonstrates the effect of the systematic errors correction and the consistency of the statistical error estimation. Specifically, the left plot in Fig. 23 shows the distributions of the time difference, δt, between the signal arrival times of two adjacent detectors before (open squares) and after (solid circles) the application of corrections of systematical timing errors described above, in comparison with the MC predictions (shadowed and no-shadowed histograms respectively). The MC predictions agree very well with the data distributions signifying that the simulation describes accurately the detector effects. It is also evident that the application of corrections removes the asymmetry in the timing distribution. Furthermore, the right plot of Fig. 23 presents the corresponding pull distribution in comparison with the MC prediction. The difference in arrival times, δt, after correcting for systematic errors, is normalized to the expected detector resolutions, $\sigma_{pair} = \sqrt{\sigma^2(T_H^1) + \sigma^2(T_H^2)}$ , where the parameterization of the timing resolution as a function of the measured ToT is used. The pull distributions, obtained from the MC events and the calibration data, follow normal distributions with mean and sigma consistent to zero and one respectively, signifying the consistency in estimating the statistical timing errors.

## 3. Evaluation of the Stations Performance

Data collected by the calibration set-up of Fig 12, were used to reconstruct the direction of the detected EAS, i.e. the zenith - $\theta$ and the azimuth - $\varphi$ of an axis normal to the incoming particle front. Experimental information (timing signals) from each Station was used independently to estimate the angular direction by applying [20] a triangulation algorithm, which is briefly described in the following.

Let $\{t_i, \vec{r}_i\} i=1,2,3$ be the arrival times of a planar particle-front at three points represented by the vectors $\vec{r}_i = (x_i, y_i, z_i)^T$, in the local Cartesian coordinate system, $XYZ$, of a Station. Consider these points to coincide with the geometrical centers of the three detectors of a HELYCON Station. Let $\vec{r}_i' = (x_i', y_i', 0)^T$ to be the coordinates of the same points at another, primed $X'Y'Z'$ frame such as that the plane $X'Y'$ is defined by the centers of the three detectors. Then

$$\vec{r}_i' = \widetilde{A} \cdot \vec{r}_i$$

$$\widetilde{A} = \begin{bmatrix} \dfrac{D_x D_z}{\sqrt{D_x^2 + D_y^2}} & \dfrac{D_y D_z}{\sqrt{D_x^2 + D_y^2}} & -\sqrt{D_x^2 + D_y^2} \\ \dfrac{-D_y}{\sqrt{D_x^2 + D_y^2}} & \dfrac{D_x}{\sqrt{D_x^2 + D_y^2}} & 0 \\ D_x & D_y & D_z \end{bmatrix} \quad (8a)$$

where the unit vector $\widehat{D} = (D_x, D_y, D_z)^T$ is defined as $\widehat{D} = \dfrac{(\vec{r}_2 - \vec{r}_1) \times (\vec{r}_3 - \vec{r}_1)}{|(\vec{r}_2 - \vec{r}_1) \times (\vec{r}_3 - \vec{r}_1)|}$

Let $\vec{d}' = (d_x', d_y', d_z')^T$ to be a unit vector that defines the direction of the particle's front, which is related to the $\{t_i, \vec{r}_i'\} i=1,2,3$ according to the following relations:

$$d_x' = c \cdot \frac{(t_1 - t_3)(y_2' - y_1') - (t_1 - t_2)(y_3' - y_1')}{(x_3' - x_1')(y_2' - y_1') - (x_2' - x_1')(y_3' - y_1')}$$

$$d_y' = c \cdot \frac{(t_1 - t_2)(x_3' - x_1') - (t_1 - t_3)(x_2' - x_1')}{(x_3' - x_1')(y_2' - y_1') - (x_2' - x_1')(y_3' - y_1')} \quad (8b)$$

$$d_z' = \sqrt{1 - d_x'^2 - d_y'^2}$$

whilst the direction of the particle's front in the local Station frame, $\widehat{d} = (d_x, d_y, d_z)^T$ derives from the inverse transformation, $\widehat{d} = \widetilde{A}^{-1} \widehat{d}'$. Finally the zenith and azimuth angles are related to the directional vector as:

$$\widehat{d} = (d_x, d_y, d_z)^T = (\sin\theta \cdot \cos\varphi, \sin\theta \cdot \sin\varphi, \cos\theta)^T \quad (8c)$$

The application of eq. (8) requires information for the arrival times of the particle's front, which is provided by the detector signals, as well the knowledge of the geometrical detectors position which have been determined by topographical methods. However, this reconstruction technique treats the shower's front as a plane, which is a good approximation at least for the part of an EAS that hits the detectors of a single Station. In principle, eq. (8) estimates the direction of move of a limited part of the particle-front (hereafter "local front"), which does not necessarily coincide with the direction of the shower's axis. Due to the fact that both Stations involved in the calibration set-up detect the same local front of an EAS, the triangulation technique should result to the same directional angles when applied to experimental data provided by one or the other of the inter-calibrated Station. The comparison of these independent estimations provides the means to study the resolution and reveal systematical biases in reconstructing the direction of particle-fronts, to check the MC accuracy in describing the performance of the HELYCON Stations as well as to evaluate the functioning of the Quarknet electronics.

This study uses two sets of calibration data, a set collected by running both the inter-calibrated Stations with digital oscilloscopes (hereafter called "uniform" data sample) and another set (hereafter called "mixed" data sample) that data were collected by running the reference Station with Quarknet electronics while the detector signals of the second Station were fully digitized by an oscilloscope as described in Section 2.1. The uniform data were accumulated as described in Section 2.4 while for the mixed data the Quarknet trigger was used, formed by a triple coincidence of the reference Station's signals when exceeding a common threshold of 4.7 mV. The events used in this analysis had to fulfill the extra offline requirement that all the detector signals of both Stations exceed in amplitude a level of 10 mV. The parameterization described in Section 2.3 was used for estimating the peak-Voltage of signals digitized by the Quarknet. This study also makes use of MC sets consisting from uniform and mixed events (where the Quarknet functions were simulated in detail) which fulfill the same selection requirement as the calibration data.

In the following analysis, the arrival time of a particle-front at the center of a counter was estimated as the time when the waveform of the corresponding signal crosses the threshold level of 4.7 mV, corrected for relative delays (due to cables etc.) and for slewing effects as described in Section 2.4. Consequently, the triangulation technique was used to estimate the directional angles, $\theta$ and $\varphi$, of a detected EAS local front, as well as their statistical uncertainties.

For each event, i.e. a triplet of measurements comprise the arrival time measurements $\{t_1^m, t_2^m, t_3^m\}$ and the corresponding ToT measurements $\{T_1^m, T_2^m, T_3^m\}$ from the three detectors of the Station, the directional angles $\{\theta^m, \varphi^m\}$ were evaluated using eq. (8) and the timing measurements. Due to the fact that the procedure described by eq. (8) is not linear, the statistical errors in estimating $\theta$ and $\varphi$, which reflect the detectors timing resolution, were evaluated by a Monte Carlo Integration method. Specifically, values of three uncorrelated random numbers, $\{t_1, t_2, t_3\}$, were selected according to the following, factorized, common pdf:

$$\prod_{i=1}^{3} P\left(t_i; t_i^m, \sigma_i\left(T_i^m\right)\right) = \prod_{i=1}^{3} \frac{1}{\sqrt{2\pi} \cdot \sigma_i\left(T_i^m\right)} \cdot \exp\left(\frac{-\left(t_i - t_i^m\right)^2}{2 \cdot \left(\sigma_i\left(T_i^m\right)\right)^2}\right) \tag{9}$$

where $i=1,2,3$ signifies the detector id and $\sigma_i\left(T_i^m\right)$ is the expected timing resolution of the $i^{th}$

detector of the Station, for a signal with ToT value (at 4.7 mV threshold level) equal to the measured $T_i^m$. As already described in the previous Section, $\sigma_i(T_i^m)$ is evaluated for all the detectors using the collected calibration data. The selected values $\{t_1, t_2, t_3\}$ were used in the triangulation procedure to reconstruct the $\theta$ and $\varphi$ directional angles. By repeating the above described step N (= 300) times, per each set of selected values $\{t_1, t_2, t_3\}$, a set of n pairs of reconstructed angles $(\{\theta_j^R, \varphi_j^R\}, j=1,2,3\ldots,n)$ are produced. In principle $n \leq N$, because not every triplet of assumed arrival times results to acceptable solutions of eq. (8). However, due to the good timing resolution of the detectors, there were very few cases with no solution.

Then, using

$$\langle \theta^R \rangle = \frac{1}{n}\sum_{j=1}^{n} \theta_j^R, \left\langle \left(\theta^R\right)^2 \right\rangle = \frac{1}{n}\sum_{j=1}^{n} \left(\theta_j^R\right)^2, \langle \varphi^R \rangle = \frac{1}{n}\sum_{j=1}^{n} \varphi_j^R, \left\langle \left(\varphi^R\right)^2 \right\rangle = \frac{1}{n}\sum_{j=1}^{n} \left(\varphi_j^R\right)^2$$

and

$$\langle \theta^R \varphi^R \rangle = \frac{1}{n}\sum_{j=1}^{n} \theta_j^R \cdot \varphi_j^R$$

the variances and covariance of the reconstructed angles were estimated as:

$$V[\theta^m] = \sigma_\theta^2 = \left\langle \left(\theta^R\right)^2 \right\rangle - \langle \theta^R \rangle^2$$
$$V[\varphi^m] = \sigma_\varphi^2 = \left\langle \left(\varphi^R\right)^2 \right\rangle - \langle \varphi^R \rangle^2$$
$$\text{cov}[\theta^m, \varphi^m] = \langle \theta^R \varphi^R \rangle - \langle \theta^R \rangle \langle \varphi^R \rangle \tag{10}$$

It should be mentioned that for the time measurements taken with the Quarknet electronics an extra equiprobable time smearing, within a window of 1.25 ns, was applied on top of the Gaussian smearing expressed by eq. (9), in order to take into account the Quarknet's time digitization step.

In general, the values of $\langle \theta^R \rangle$ and $\langle \varphi^R \rangle$ are different but very close to the values $\theta^m$ and $\varphi^m$, which have been chosen as estimators of the local particle-front's direction. As it is also expected, the correlation between the estimations $\theta^m$ and $\varphi^m$, expressed as $\rho = \text{cov}[\theta^m, \varphi^m]/\sigma_\theta \cdot \sigma_\varphi$, is not negligible; in the analyzed events it is found that ρ is distributed symmetrically around zero with an RMS of 35%.

Fig. 24 demonstrates the resolution of HELYCON Stations in reconstructing the directional angles of EAS local fronts. The left plot presents the distribution of $\Delta\theta = \theta_1^m - \theta_2^m$, between the θ-angles of the same EAS which are reconstructed using experimental information from the reference Station 1 $(\theta_1^m)$ and from the other Station $(\theta_2^m)$. The right plot presents the corresponding difference in $\Delta\varphi$. In both plots, the solid circles are related to the mixed whilst the open squares to the uniform calibration data sets. The histograms represent MC predictions for mixed data samples. All the shown distributions are symmetrical with mean values consistent with zero, demonstrating a good agreement of the MC predictions with the calibration data. Furthermore, the very good agreement between the distributions of the mixed and the uniform data sets demonstrates that the HELYCON Stations performed the same way when running with fast oscilloscopes or Quarknet electronics. The resolution of a typical HELYCON station in reconstructing the zenith angle of the local particle-front, quantified by the $RMS/\sqrt{2}$ of the $\Delta\theta$ distribution, is found to be 2.9°, whilst the resolution in $\varphi$ found to be 8.5°.

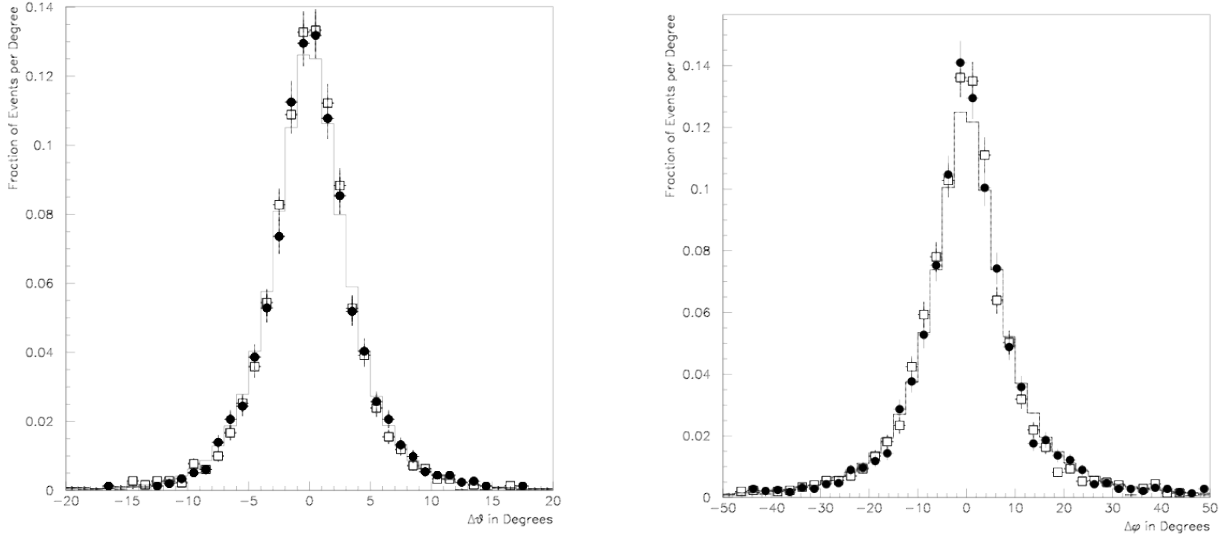

Figure 24: Demonstration of the HELYCON Stations resolution in reconstructing the directional angles of EAS local fronts. The left plot presents the distribution of the difference in estimated zenith angles ($\theta$) when the same EAS is reconstructed by the reference Station 1 and the Station 2. The right plot shows the distribution of the corresponding difference in azimuth angle ($\varphi$). In both plots, the solid circles are related to the mixed whilst the open squares to the uniform calibration data sets. The histograms represent MC predictions for mixed data samples.

Due to the non zero correlation in the $\theta$ and $\varphi$ estimations by a single Station, the pull distribution is not an adequate check of the consistency in estimating the statistical errors described above. Instead, the flatness of $\chi^2$ distribution was used as a qualification check. The reconstructed $\theta_k^m$ and $\varphi_k^m$ angles by the $k=1,2$ Station of the calibration set-up (estimated with $\sigma_{\theta,k}, \sigma_{\varphi,k}, C_k = cov[\theta_k^m, \varphi_k^m]$ statistical errors and covariance, respectively), can be viewed as two measurements of the same, "true" angles $\theta$ and $\varphi$, where correlation terms between measurements by different detectors are zero. A combination of the above measurements can be performed by a $\chi^2$ minimization with two degrees of freedom, i.e four measurements $(\theta_1^m, \theta_2^m, \varphi_1^m, \varphi_2^m)$ and two estimated values ($\theta$, $\varphi$), which results to the minimum chi squared $(\chi^2_{min})$ value given by the following expression:

$$\chi^2_{min} = \frac{1}{1-\rho^2} \cdot \left[ \frac{(\theta_1^m - \theta_2^m)^2}{\sigma_{\theta_1}^2 + \sigma_{\theta_2}^2} + \frac{(\varphi_1^m - \varphi_2^m)^2}{\sigma_{\varphi_1}^2 + \sigma_{\varphi_2}^2} - 2\rho \frac{(\theta_1^m - \theta_2^m) \cdot (\varphi_1^m - \varphi_2^m)}{\sqrt{(\sigma_{\theta_1}^2 + \sigma_{\theta_2}^2) \cdot (\sigma_{\varphi_1}^2 + \sigma_{\varphi_2}^2)}} \right] \quad (11)$$

where

$$\rho = \frac{C_1 + C_2}{\sqrt{(\sigma_{\theta_1}^2 + \sigma_{\theta_2}^2) \cdot (\sigma_{\varphi_1}^2 + \sigma_{\varphi_2}^2)}}$$

In case of an unbiased estimation, i.e. the estimated direction angles $\left(\theta_1^m, \theta_2^m, \varphi_1^m, \varphi_2^m\right)$ and their covariant elements are unbiased, the $\chi^2$-probability for two degrees of freedom of the $\chi^2_{min}$ should be distributed uniformly between zero and one.

As it shown in Fig. 25, the probability of the $\chi^2_{min}$ for the vast majority of analyzed events it is indeed flat, with the exception of a less than 10% fraction of events that exhibit low probability values, signifying «bad reconstructions». The same plot demonstrates the very good agreement of the MC prediction with calibration data collected by both DAQ configurations, i.e. when the data have been digitized using either the Quarknet electronics or the fast oscilloscope.

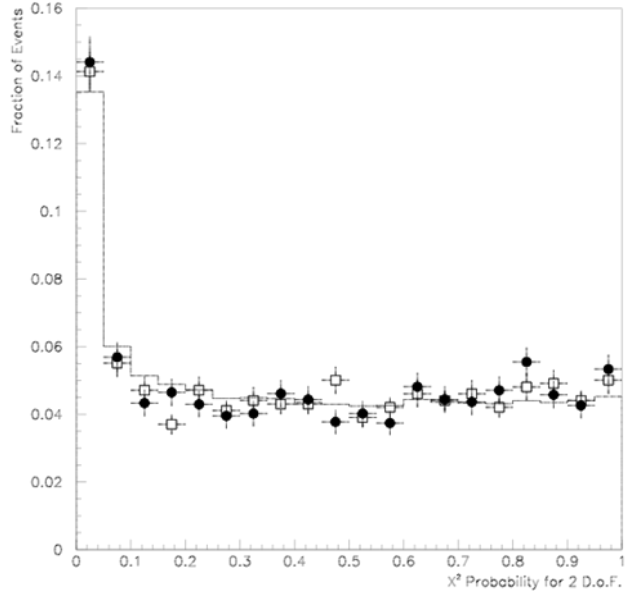

Figure 25: The $\chi^2$-probability for two degrees of freedom of the $\chi^2_{min}$ defined as in eq. (11). The histogram represents the MC prediction whilst the solid circles and the open squares depict the estimation performance of using the mixed and the uniform data sets respectively.

Due to the good agreement between MC and calibration data $\chi^2$-probability distributions, the simulated events have been used to study the causes of the bad reconstructions. It was found that these events correspond to cases where the impact point of EAS's axis to the ground is far from the geometrical center of the HELYCON station. Specifically, in these cases the average radial distance of the impact point from the Station's center found to be about 400 m, in comparison with 200 m average distance for the rest of the events. In these cases, the HELYCON counters detect particles at the fringes of the EAS particles front, where the basic reconstruction assumption for the flatness of the local front is not anymore accurate. Furthermore, the physical particle-density fluctuations of the particles front cause fluctuations to the relative arrival times (see eq. (8b)). These extra timing fluctuations are not included in the timing errors parameterization, described in Section 2.4, which is used to evaluate the statistical errors ( $\sigma_{\theta,k}, \sigma_{\varphi,k}, C_k = cov\left[\theta_k^m, \varphi_k^m\right]$ ) in estimating the directional angles. Consequently, in these cases, the errors used in eq. (11) are underestimated resulting in higher $\chi^2_{min}$ values and in the excess of entries at low $\chi^2$-probability values shown in Fig. 25.

Finally, Fig. 26 presents the 3d angular difference. There is a very good agreement between the MC prediction with the calibration data, which were accumulated either by the oscilloscope or the

Quarknet electronics. The mean 3d angle between the two estimations is 4.7° whilst the median is 3.5°. However, using the estimated events, the 3d angular difference of the HELYCON Station directional reconstruction to the «true» EAS axis, is found to have a mean of 4° whilst the median is 3°.

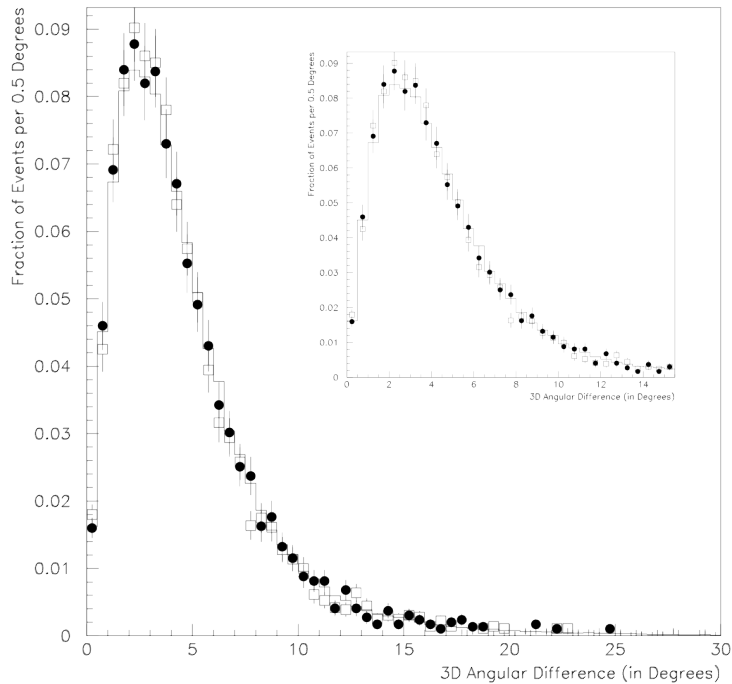

Figure 26: The angular difference between the two reconstructed axis of the local particle-front by the two intercalibrated Stations. The histogram represents the MC prediction whilst the solid circles and open squares depict the calibration data collected by the Quarknet and oscilloscope DAQs respectively.

## 4. Conclusions

We have deployed three HELYCON stations (each one including three scintillator detectors and one CODALEMA antenna) at the Hellenic Open University campus. Each station is autonomous, bearing its own control and monitor electronics and DAQ system and operates continuously. Remote control of the operation of the stations and real time access to the data is available. Standard procedures for the installation, calibration and operation of the stations have been established and software for the remote control, the real time monitor and the stations' data collection has been developed. Software packages have also been developed for the detailed simulation of the detectors response to particle showers and for the reconstruction of the EAS characteristics. Each station has been calibrated and the simulation package parameters have been fine tuned. The stations performance in reconstructing the angular direction of the EAS particle-front has been evaluated and the results have been found in very good agreement with the predictions of the simulation.

The detector network has been operated, collecting data for more than a year and a half. The analysis of the data collected from each station, as well as of the data from more than one station that correspond to the same EAS, and the analysis of data recorded synchronously by both the large particle detectors and the CODALEMA RF antennas of a HELYCON station, will be published in a following paper.

## 5. Acknowledgments

This research has been co-financed by the European Union (European Social Fund – ESF) and Greek national funds through the Operational Program "Education and Lifelong Learning" of the National Strategic Reference Framework (NSRF) – Research Funding Program: "THALIS – Hellenic Open University – Development and Applications of Novel Instrumentation and Experimental Methods in Astroparticle Physics".

## References

[1] S.E. Tzamarias, HELYCON: towards a sea-top infrastructure, in: Proceedings of the 6th International Workshop on the Identification of Dark Matter (IDM 2006), World Scientific (2007), p. 464 (ISBN-13978-981-270-852-6).

[2] D. Ardouin A. Belltoile D. Charrier et al. Radio-detection signature of high-energy cosmic rays by the CODALEMA experiment. Nucl. Instrum. Methods A, 555:148163, 2005.

[3] A. Leisos, G. Bourlis. A.G. Tsirigotis, and S.E. Tzamarias on behalf of the KM3NeT consortium. Synchronous Detection of Extensive Air Showers by a HELYCON Detector Array and a Deep Sea Underwater Neutrino Telescope: Statistical and Systematic Effects, Nucl. Instr. and Meth. A

[4] A.Leisos, G. Bourlis, P.E. Christopoulou, A. G. Tsirigotis, S.E.Tzamarias. Calibration and Optimization of a Very Large Volume Neutrino Telescope using Extensive Air Showers, to appear in Nucl. Instr. and Meth. A.

[5] A. G. Tsirigotis et al. Use of floating surface detector stations for the calibration of a deep-sea neutrino telescope, 2008 (Cited 1 time) (http://dx.doi.org/10.1016/j.nima.2008.07.011)

[6] George Bourlis, Ph.D. Thesis, "Development of instrumentation and methodology for the detection of atmospheric cosmic ray showers and applications in calibrating an underwater neutrino telescope", http://physicslab.eap.gr/EN/Publications_files/Bourlis_110430.pdf

[7] IHEP Scintillators http://www.fnal.gov/projects/ckm/photon_veto/scint/ihep_plastic_production/catalogue/moldscint.html

[8] Bicron BCF91-A Optical Fibers http://www.crystals.saint-gobain.com/sites/imdf.crystals.com/files/documents/fiber-brochure.pdf

[9] XP1912 Photomultiplier Tubes Product Specifications. http://hzcphotonics.com/products/xp1912.pdf.

[10] EMCO High Voltage, CA Series Precision Regulated, Low Ripple High Voltage DC to DC Converters, http://www.emcohighvoltage.com/regulated/caseries.php

[11] National Instruments NI USB-6008, 12-bit, 10KS/s multifunction DAQ, http://sine.ni.com/nips/cds/view/p/lang/en/nid/201986

[12] J.Rylander T.Jordan J.Paschke H.-G.Berns. Quarknet Cosmic Ray Myon Detector User's Manual Series \6000" DAQ. Fermilab, Univ. of Nebraska, Univ. of Washington, January 2010.

[13] A. Tsirigotis, "HELYCON: A Status Report", Proceedings of the 20th European Cosmic Ray Symposium

[14] Theodore Avgitas, Master Thesis, "Atmospheric Showers of Energetic Cosmic Particles: Detection and Reconstruction", http://nemertes.lis.upatras.gr/jspui/handle/10889/7903

[15] CAEN Mod. N978 Variable Gain Fast Amplifier http://www.caen.it/csite/CaenProd.jsp?parent=12&idmod=440

[16] G. Bourlis, T. Avgitas, A. Leisos, I. Manthos, A.G. Tsirigotis, S.E. Tzamarias, A Data Acquisition System based on high sampling rate oscilloscopes, PCI2016, Patras, Greece, 10-12 Nov. 2016.

[17] A.G. Tsirigotis A. Leisos S.E. Tzamarias. HOU reconstruction & simulation (HOURS): A complete simulation and reconstruction package for very large volume underwater neutrino telescopes. Nuclear Instruments and Methods in Physics Research Section A: Accelerators, Spectrometers, Detectors and Associated Equipment, 626627:185-187, 2011.

[18] D. Heck, J. Knapp, J. N. Capdevielle, G. Schatz, and T. Thouw. CORSIKA: a Monte Carlo code to simulate extensive air showers. Forschungszentrum Karlsruhe GmbH, Karlsruhe (Germany), Feb 1998, V + 90 p., TIB Hannover, D-30167 Hannover (Germany), Feb 1998.

[19] B. Wiebel-Sooth P. L. Biermann. Cosmic Rays. Springer Verlag, 1998

[20] T. Avgitas, G. Bourlis, G.K. Fanourakis, I. Gkialas, A. Leisos, I. Manthos, A. Tsirigotis, S.E. Tzamarias, Operation of a pilot HELYCON cosmic ray telescope with 3 stations, to be published.